\documentclass[11pt,onecolumn,draftcls]{IEEEtran} 
\usepackage{graphicx,psfrag,amsthm,subfig,fancyvrb}
\usepackage[usenames]{color}
\usepackage[cmex10]{amsmath}
\usepackage{dsfont}
\usepackage{bbm}
\usepackage{array}
\usepackage{amsfonts,amssymb,latexsym,cite,ifthen,cite,hyperref}
\newboolean{SEP_FIG_CAPS}\setboolean{SEP_FIG_CAPS}{false}

\ifthenelse{\boolean{SEP_FIG_CAPS}}
{

 \newcommand{\capFrag}[2]{\noindent Fig.~\ref{fig:#1}. #2 \medskip\\}
 \newcommand{\capTable}[2]{\noindent Tab.~\ref{tab:#1}. #2 \medskip\\}
}
{

 \newcommand{\capFrag}[2]{}
 \newcommand{\capTable}[2]{}
}

 \newcommand{\defn}{\triangleq}

 \newcommand{\tvec}[1]{\ensuremath{\Tilde{\boldsymbol{#1}}}}

 \renewcommand{\vec}[1]{\ensuremath{\boldsymbol{#1}}}

 \newcommand{\mc}[1]{\ensuremath{\mathcal{#1}}}

 \newcommand{\Complex}{{\mathbb{C}}}

 \DeclareMathOperator{\E}{E}
 
 \DeclareMathOperator{\cov}{Cov}

 \DeclareMathOperator*{\argmax}{argmax}
 \DeclareMathOperator*{\argmin}{argmin}

 \newtheorem{lemma}{Lemma}

 \renewcommand{\eqref}[1]{(\ref{eq:#1})}
 
 \newcommand{\Figref}[1]{Figure~\ref{fig:#1}}
 \newcommand{\figref}[1]{Fig.~\ref{fig:#1}}
 
 \newcommand{\secref}[1]{Section~\ref{sec:#1}}

 \newcommand{\lemref}[1]{Lemma~\ref{lem:#1}}

 \newcommand{\ie}{i.e., }

 \newcounter{comment}[section]
 
 \newcounter{texthead}[section]

 \newcommand{\dsra}{_{\text{\sf DSRA}}}
 \newcommand{\csra}{_{\text{\sf CSRA}}}

 \newcommand{\overall}{_{\text{\sf tot}}}
 \newcommand{\con}{_{\text{\sf con}}}

\begin{document}
\setlength{\arraycolsep}{0.8mm}
 \title{\huge 
Joint Scheduling and Resource Allocation in the OFDMA Downlink: Utility Maximization under Imperfect Channel-State Information}
	 \author{
	 \IEEEauthorblockN{
	 Rohit Aggarwal\IEEEauthorrefmark{1},
	 		Mohamad Assaad\IEEEauthorrefmark{2},
	 		C.~Emre Koksal\IEEEauthorrefmark{1}, and 
			Philip Schniter\IEEEauthorrefmark{1}}	\\		
         \IEEEauthorblockA{\IEEEauthorrefmark{1}Dept. of ECE,
	 		   The Ohio State University,
			   Columbus, OH 43210. \\
			   Email: \{aggarwar,koksal,schniter\}@ece.osu.edu} \\
	 \IEEEauthorblockA{\IEEEauthorrefmark{2}Dept. of Telecom., Sup{\'e}lec, France.\\
			   Email: mohamad.assaad@supelec.fr}
	}
 \date{\today}
 \maketitle

\begin{abstract}
We consider the problem of simultaneous user-scheduling, power-allocation, 
and rate-selection in an OFDMA downlink, 
with the goal of maximizing expected sum-utility under a sum-power constraint. 
In doing so, we consider a family of generic goodput-based utilities that facilitate, 
e.g., throughput-based pricing, quality-of-service enforcement, and/or the treatment 
of practical modulation-and-coding schemes (MCS).
Since perfect knowledge of channel state information (CSI) may be difficult to 
maintain at the base-station,
especially when the number of users 
and/or subchannels is large, we consider scheduling and resource allocation 
under imperfect CSI, where the channel state
is described by a generic probability distribution.
First, we consider the ``continuous'' case where multiple users and/or code rates
can time-share a single OFDMA subchannel and time slot.
This yields a non-convex optimization problem that we convert into a convex 
optimization problem and solve exactly using a dual optimization approach. 
Second, we consider the ``discrete'' case where 
only a single user and code rate is allowed per OFDMA subchannel per time slot.
For the mixed-integer optimization problem that arises, we discuss the
connections it has with the continuous case and show that it can
solved exactly in some situations. 
For the other situations, we present a bound on the optimality gap. 
For both cases, we provide algorithmic implementations of the obtained solution. 
Finally, we study, numerically, the performance of the proposed algorithms
under various degrees of CSI uncertainty, utilities, and OFDMA system configurations.
In addition, we demonstrate advantages relative to existing state-of-the-art algorithms.

\end{abstract}

\section{Introduction} 				\label{sec:intro}
In the downlink of a wireless orthogonal frequency division multiple 
access (OFDMA) system, the base station (BS) delivers data to a pool 
of users whose channels vary in both time and frequency. 
Since bandwidth and power resources are limited, the BS would like to 
allocate them most effectively, e.g., by pairing users with strong 
subchannels and distributing power 
{%
in order to maximize some function of the delivered data rates \cite{Song3}.
Although, for resource allocation, one would ideally like to have access to
instantaneous channel state information (CSI), such CSI is difficult to 
obtain in practice, and so resource allocation must be accomplished under 
imperfect CSI.
Thus, in this paper, we consider simultaneous user-scheduling, power-allocation, 
and rate-selection in an OFDMA downlink, given only a generic \emph{distribution} 
for the subchannel signal-to-noise ratios (SNRs), with the goal of maximizing 
expected sum-utility under a sum-power constraint. 
In doing so, we consider relatively generic goodput-based utilities, facilitating, 
e.g., throughput-based pricing (e.g., \cite{Pricingsurvey, Pricingsurvey2,shenker}), 
quality-of-service enforcement, and/or the treatment of practical modulation-and-coding 
schemes (MCS).
}

{%
In particular, we consider the above scheduling and resource allocation (SRA) problem 
under two scenarios.
In the first scenario, we allow multiple users (and/or MCSs) to time-share 
any given subchannel and time-slot. 
In practice, this scenario occurs, e.g., in OFDMA systems where several users are 
multiplexed within a time-slot, such as IEEE $802.16/\textrm{WiMAX}$~\cite{wimax} 
and 3GPP LTE~\cite{LTE2}. 
Although the resulting optimization problem is non-convex, we
show that it can be converted into a convex problem and 
solved exactly using a dual optimization approach. 
Based on a detailed analysis of the optimal solution, we propose a novel
bisection-based algorithm that is faster than state-of-the-art 
golden-section based approaches (e.g., 
\cite{berry:tcom:2009}) and that admits finite-iteration 
performance guarantees.
} 
In the second scenario, we allow at most one combination of user and 
MCS to be used on any given subchannel and time-slot. 
{%
This scenario occurs widely in practice, such as in the Dedicated Traffic 
Channel (DTCH) mode of UMTS-LTE \cite{UMTS}, and results in a  
mixed-integer optimization problem.
Based on a detailed analysis of the optimal solution to this problem  
and its relationship to that in the first scenario, we propose a novel
suboptimal algorithm 
that is faster than state-of-the-art golden-section and subgradient
based approaches (e.g., \cite{berry:tcom:2009,TCOM:wong:09}), 
and we derive a novel tight bound on the optimality gap of our algorithm.
Finally, we simulate our algorithms under various OFDMA system configurations,
comparing against state-of-the-art approaches and genie-aided performance bounds.
}

{%
We now discuss related work.
The problem of OFDMA downlink SRA 
under \emph{perfect} CSI has been studied in several papers, notably
\cite{Song1,Oct:JSAC:Wong:99,
Mar:TIT:Willink:97,Jun:TCOM:Hoo:04,Mar:TWC:Wong:08,Jul:PIIT:Seong:06}. 
In \cite{Song1},
a utility maximization framework for discrete allocation was formulated 
to balance system efficiency and fairness, and efficient subgradient-based 
algorithms were proposed. 
In \cite{Oct:JSAC:Wong:99}, a subchannel, rate, and 
power allocation algorithm was developed to minimize power consumption 
while maintaining a total rate-allocation requirement for every user.
In~\cite{Mar:TWC:Wong:08}, a weighted-sum capacity maximization problem
with/without subchannel sharing was formulated to allocate subcarriers and powers. 
In \cite{Jul:PIIT:Seong:06}, non-convex optimization problems regarding weighted 
sum-rate maximization and weighted sum-power minimization were solved using a Lagrange 
dual decomposition method.
Compared to the above works, we extend the utility
maximization framework to imperfect CSI and continuous allocations,
and propose bisection-based algorithms that are faster for both the discrete and 
continuous allocation scenarios. 
Unlike \cite{Song1,Oct:JSAC:Wong:99, 
Mar:TIT:Willink:97,Jun:TCOM:Hoo:04,Mar:TWC:Wong:08,Jul:PIIT:Seong:06},
our utility framework can be applied to problems with/without fixed rate-power 
functions\footnote{By a ``fixed rate-power function'' we mean that, for a given SNR, 
the achievable rate is a known function of the power.}. 
In additional, it can be applied to pricing-based utilities (e.g., responsive pricing 
and proportional fairness pricing)~\cite{Pricingsurvey}. 
Furthermore, we study the relationship between the discrete and continuous allocation 
scenarios, and provide a tight bound on the duality gap of our proposed 
discrete-allocation scheme.}

{%
The problem of OFDMA downlink SRA under \emph{imperfect} CSI was studied in several
papers, notably
\cite{TCOM:wong:09,berry:tcom:2009,PIMRC:ahmad:09,July:TWC:Hui:09}. 
In \cite{TCOM:wong:09},
the authors considered the problem of discrete ergodic weighted sum-rate
maximization for user scheduling and resource allocation, and studied the impact 
of channel estimation error due to pilot-aided MMSE channel estimation. 
In \cite{berry:tcom:2009}, a deterministic optimization problem was 
formulated using an upper bound on system capacity (via Jensen's inequality) 
as the objective. 
Both optimal and heuristic algorithms were then proposed to implement the obtained solution.
Compared to these two works, we propose faster algorithms, applicable to 
a general utility maximization framework (of which the objectives 
in~\cite{TCOM:wong:09,berry:tcom:2009} are special cases), 
under a more general class of channel estimators, and for both discrete and continuous
subchannel allocations. 
Our algorithms are inspired by a rigorous analysis of the optimal solutions to the 
discrete and continuous problems.
In \cite{PIMRC:ahmad:09}, the problem of total transmit 
power minimization, subject to strict constraints on conditional expected 
user capacities, was investigated. 
In~\cite{July:TWC:Hui:09}, the
effect of heterogeneous delay requirements and outdated CSI on a particular
discrete resource allocation problem was studied. 
In contrast, we consider a general utility maximization problem that 
allows us to attack problems that may or may not be based on fixed rate-power 
functions, as well as those based on pricing models. 
Relative to these works, we propose faster algorithms for both continuous and 
discrete allocation problems with provable bounds on their performances.}

The remainder of this paper is organized as follows. 
In \secref{model}, we outline the system model and frame our optimization 
problems. 
In \secref{continuous}, we consider the ``continuous'' problem,
where each subchannel can be shared by multiple users and rates, 
and find its exact solution.
In \secref{discrete}, we consider the ``discrete'' problem,
where each subchannel can support at most one combination of user and rate
per time slot. 
In \secref{simulations}, we compare the performance of the 
proposed algorithms to reference algorithms under various settings.
Finally, in \secref{conclusion}, we conclude.

\section{System Model}					\label{sec:model}

We consider a downlink OFDMA system with $N$ subchannels and 
$K$ active users ($N,K\in\mathbb{Z}^+$) as shown in \figref{system-model}. 
The scheduler-and-resource-allocator at the base-station uses the 
imperfect CSI to send data to the users, across 
OFDMA subchannels, in a way that maximizes utility.
We assume that, for each user, there is an infinite backlog of data at 
the base-station, so that there is always data available to be transmitted. 
{
During every channel use and across every OFDMA subchannel,
the base-station transmits codeword(s) from a generic signaling scheme, 
which propagate to the intended mobile recipient(s)  
through their respective fading channels.}
For a given user $k$, the OFDMA subchannels are assumed to be non-interfering,
with gains that are time-invariant over each codeword duration
{%
and statistically independent of those for other users.}
Thus, the successful reception of a transmitted codeword depends on the 
corresponding subchannel's SNR $\gamma$, power $p$, and modulation and coding 
scheme (MCS), indexed by $m\in\{1, \ldots, M\}$. 
We assume that, for user $k$, MCS $m$ corresponds 
to a transmission rate of $r_{k,m}$ bits per codeword and a codeword error 
probability of 
{%
$\epsilon_{k,m}(p \gamma) = a_{k,m} e^{-b_{k,m} p \gamma}$ 
for known constants $a_{k,m}$ and $b_{k,m}$ (see, e.g., \cite{TCOM:wong:09}). 
Here, the subchannel SNR $\gamma$ is treated as an exogenous parameter, 
so that $p \gamma$ is the effective received SNR.}

To precisely state our scheduling and resource allocation 
(SRA) problem, some additional notation is useful.
To indicate how subchannels are partitioned among users and 
rates in each time-slot, we will use the proportionality indicator 
$I_{n,k,m}$, where $I_{n,k,m}=1$ means that subchannel $n$ is fully 
dedicated to user $k$ at MCS $m$, and $I_{n,k,m}=0$ means that 
subchannel $n$ is totally unavailable to user $k$ at MCS $m$. The 
subchannel resource constraint is then expressed as 
$\sum_{k,m} I_{n,k,m} \leq 1$ for all $n$. 
In the sequel, we 
consider two flavors of the SRA problem, a ``continuous'' one 
where each subchannel can be shared among multiple users and/or 
rates per time slot (\ie $I_{n,k,m}\in[0,1]$), and a ``discrete'' 
one where each subchannel can be allocated to at most one 
user/rate combination per time slot (\ie $I_{n,k,m}\in\{0,1\}$). 
We will use $p_{n,k,m}\geq 0$ as the power that would be 
expended on subchannel $n$ if it was fully allocated to the 
user/rate combination $(k,m)$. 
With this definition, the total 
expended power becomes $\sum_{n,k,m} I_{n,k,m} p_{n,k,m}$. 
Finally, we will use $\gamma_{n,k}$ to denote the $n^{th}$ subchannel's 
SNR for user $k$. 
Although we assume that the BS does not know the SNR 
realizations $\{\gamma_{n,k}\}$, we assume that it does know the 
(marginal) distribution of each $\gamma_{n,k}$.

{%
When subchannel $n$ is fully dedicated to user $k$ with MCS $m$ and power $p_{n,k,m}$, 
the goodput $g_{n,k,m} = (1-a_{k,m} e^{-b_{k,m} p_{n,k,m} \gamma_{n,k}})r_{k,m}$ 
quantifies the expected number of bits, per codeword, transmitted without 
error.
In the sequel, we focus on maximizing goodput-based utilities 
of the form $U_{n,k,m}(g_{n,k,m})$, where $U_{n,k,m}(\cdot)$ is any generic real-valued
function that is twice differentiable, strictly-increasing, and concave,
with $U_{n,k,m}(0) < \infty$. 
(These conditions imply $U'_{n,k,m}(\cdot) > 0$ and $U''_{n,k,m}(\cdot) \leq 0$.)
In particular, we aim to maximize the expected sum utility,
$\E\big\{\sum_{n,k,m} I_{n,k,m} U_{n,k,m}(g_{n,k,m})\big\}$,
where the expectation is taken over the subchannel-SNRs $\{\gamma_{n,k}\}$ 
hidden within the goodputs.
} 
Incorporating a sum-power constraint of $P\con$, our SRA problem becomes
\begin{eqnarray}
\text{SRA} 
&\defn& \max_{\substack{\scriptstyle \{p_{n,k,m} \geq 0\} \\[0.8 mm]
	\scriptstyle \{I_{n,k,m}\}}} 
	\E\left\{\sum_{n=1}^N \sum_{k=1}^K \sum_{m=1}^M I_{n,k,m} U_{n,k,m}\big( 
	(1-a_{k,m} e^{-b_{k,m} p_{n,k,m} \gamma_{n,k}})r_{k,m}\big) 
	\right\} \label{eq:SRA}\\
&&~\text{s.t.}~~ \sum_{k,m} I_{n,k,m} \leq 1~\forall n ~~\text{and}~ 
 \sum_{n,k,m} I_{n,k,m} p_{n,k,m} \leq P\con \nonumber .
\end{eqnarray}
{%
The above formulation is sufficiently general to address a wide class of objectives. 
For example, to maximize sum-goodput, one would simply use $U_{n,k,m}(g)=g$.
For weighted sum-goodput, one would instead choose 
$U_{n,k,m}(g)= w_k g$ with appropriately chosen weights $\{w_k\}$.
To maximize weighted sum capacity 
$\sum_{n,k}w_k I_{n,k,1} \log \big(1+p_{n,k,1}\gamma_{n,k}\big)$, 
as in \cite{TCOM:wong:09}, one would choose $M=a_{k,1} = b_{k,1} = r_{k,1} = 1$, 
and set $U_{n,k,1}(g) = w_k \log \big(1 - \log(1- g)\big)$ for $g\in [0, 1)$. 
Commonly used utilities constructed from concave functions of capacity
$\log(1+p_{n,k,1}\gamma_{n,k})$, such as
max-min fairness and the utilities in~\cite{Song1} and \cite{berry:tcom:2009},
can also be handled by our formulation. 
For example, the utility $U_{n,k,m}(g) = 1-e^{-w_k g}$ 
(for some positive $\{w_k\}$) 
is appropriate for ``elastic'' applications such as file transfer~\cite{Pricingsurvey2,shenker}.
Our formulation also supports various pricing 
models~\cite{Pricingsurvey}, such as flat-pricing, responsive pricing, proportional 
fairness pricing, and effective-bandwidth pricing.
}

{
Next, in \secref{continuous}, we study the SRA problem for the continuous
case $I_{n,k,m} \in [0, 1]$, and in \secref{discrete} we study it 
for the discrete case $I_{n,k,m} \in \{0, 1\}$.
}

\section{Optimal Scheduling and Resource Allocation with subchannel sharing}

\label{sec:continuous}

In this section, we address the SRA problem in the case where 
$I_{n,k,m} \in [0, 1]~\forall (n,k,m)$. Recall that this problem 
arises when sharing of any subchannel by multiple users and/or 
multiple MCS combinations is allowed. We refer to this problem 
as the ``continuous scheduling and resource allocation'' (CSRA) 
problem. Defining $\vec{I}$ as the $N \times K \times M$ matrix 
with $(n,k,m)^{\textrm{th}}$ element as $I_{n,k,m}$ and the domain
of $\vec{I}$ as
\begin{equation}
\textstyle
\mathcal{I}\csra := \big\{\vec{I} : \vec{I} \in [0,1]^{N \times K \times 
M},~ \sum_{k,m}I_{n,k,m} \leq 1 ~\forall n\big\}, \nonumber
\end{equation}
the CSRA problem can be stated as
\begin{eqnarray}
\text{CSRA}	 &:=& \hspace{-2mm}
	\min_{\substack{\scriptstyle \{p_{n,k,m} \geq 0\}\\[0.8 mm]
	\scriptstyle \vec{I} \in \mathcal{I}\csra}} 
	- \hspace{-1mm}\sum_{n,k,m}\hspace{-1mm} I_{n,k,m}
	\E\Big\{ U_{n,k,m}\big((1-a_{k,m} e^{-b_{k,m} p_{n,k,m} \gamma_{n,k}})r_{k,m}\big) 
	\Big\} 
	~\text{s.t.}\sum_{n,k,m} I_{n,k,m}\,p_{n,k,m} \leq P\con. 
	\nonumber\\[-7mm]\label{eq:csra}
\end{eqnarray}
This problem has a non-convex constraint set, making it a non-convex 
optimization problem.
In order to convert it into a convex optimization problem, we write the 
``actual'' power allocated to user $k$ at MCS $m$ on subchannel $n$ as
$x_{n,k,m} = I_{n,k,m}\,p_{n,k,m}$. Then, the problem becomes
\begin{eqnarray}
\text{CSRA}	 &=& \min_{\substack{\scriptstyle \{x_{n,k,m} \geq 0\}\\[0.8 mm]
	\scriptstyle \vec{I} \in 
	\mathcal{I}\csra}} \sum_{n,k,m} I_{n,k,m}\,F_{n,k,m}(I_{n,k,m},x_{n,k,m}) ~~~\text{s.t.}~\sum_{n,k,m} x_{n,k,m} \leq P\con,
\label{eq:CSRAP}
\end{eqnarray}
where $F_{n,k,m}(\cdot,\cdot)$ is given by
\vspace{-2mm}
\begin{equation}
F_{n,k,m}(I_{n,k,m},x_{n,k,m}) = {\begin{cases}
	- \E\Big\{ U_{n,k,m}\big((1-a_{k,m} e^{-b_{k,m} x_{n,k,m} \gamma_{n,k}/I_{n,k,m}})
	r_{k,m}\big) \Big\} & \textrm{if}~ I_{n,k,m} \neq 0 \\
0 & \textrm{otherwise}. \label{eq:defF}
\end{cases}}
\end{equation}
The modified problem in~\eqref{CSRAP} is a convex
optimization problem with a convex objective function 
and linear inequality constraint. 
Moreover, Slater's condition is satisfied at $I_{n,k,m} = \frac{1}{2KM}$
and $x_{n,k,m} = \frac{P\con}{N}I_{n,k,m},~\forall n,k,m$.
Hence, the solution of \eqref{CSRAP} is the same as that of its 
dual problem (\ie zero duality gap) \cite{boyd}. Let us denote the 
optimal $\vec{I}$ and $\vec{x}$ for \eqref{CSRAP} by $\vec{I}^*\csra$
and $\vec{x}^*\csra$, respectively, and let $\vec{p}^*\csra$ be the 
corresponding $\vec{p}$.

Writing the dual formulation, using $\mu$ as the dual variable, the 
Lagrangian of \eqref{CSRAP} is
\begin{equation}
L(\mu,\vec{I},\vec{x}) = \sum_{n,k,m}  I_{n,k,m}\,F_{n,k,m}(I_{n,k,m},x_{n,k,m}) + 
\Big(\sum_{n,k,m}x_{n,k,m} - P\con\Big) \mu,  \label{eq:clagrange}
\end{equation}
where we use $\vec{x}$ to denote the $N \times K \times M$ matrix $[x_{n,k,m}]$.
The corresponding unconstrained dual problem, then, becomes
\vspace{-2mm}
\begin{eqnarray}
\lefteqn{
\max_{\scriptstyle \mu \geq 0}
\min_{\substack{\scriptstyle \vec{x}\succeq 0 \\[0.8 mm]
	\scriptstyle \vec{I}\in \mc{I}\csra}}
	L(\mu,\vec{I},\vec{x})
}\label{eq:cdual}\\
&=& \max_{\scriptstyle \mu \geq 0}
	\min_{\substack{\scriptstyle \vec{I}\in \mc{I}\csra}}
	L(\mu,\vec{I},\vec{x}^*(\mu, \vec{I}))
= \max_{\scriptstyle \mu \geq 0}
	L(\mu,\vec{I}^*(\mu),\vec{x}^*(\mu, \vec{I}^*(\mu))) 
= L(\mu^*,\vec{I}^*(\mu^*),\vec{x}^*(\mu^*, \vec{I}^*(\mu^*))),
\quad	\nonumber
\end{eqnarray}
where $\vec{x}\succeq 0$ means that $x_{n,k,m} \geq 0~\forall n,k,m$, 
$\vec{x}^*(\mu, \vec{I})$ denotes the optimal $\vec{x}$ for a given $\mu$
and $\vec{I}$, $\vec{I}^*(\mu) \in \mc{I}\csra$ denotes the optimal 
$\vec{I}$ for a given $\mu$, and $\mu^*$ denotes the optimal $\mu$. 

In the next few subsections, we will 
optimize the Lagrangian according to \eqref{cdual} w.r.t. $\vec{x}$, $\vec{I}$, 
and $\mu$ in \secref{continuous p}, \secref{continuous I}, and 
\secref{continuous mu}, respectively. We then propose an
iterative algorithm to solve CSRA problem in \secref{continuous imple}. 
Finally, we discuss some important properties of the CSRA solution in
\secref{continuous prop}. 

\subsection{Optimizing over total powers, $\vec{x}$, for a given $\mu$ 
and user-MCS allocation matrix $\vec{I}$} \label{sec:continuous p}

The Lagrangian in \eqref{clagrange} is a convex function of 
$\vec{x}$. Therefore, any local minimum of the function is a global 
minimum. Calculating the derivative of $L(\mu,\vec{I},\vec{x})$ 
w.r.t. $x_{n,k,m}$, we get
\begin{eqnarray}
\frac{\partial L(\mu,\vec{I},\vec{x})}{\partial x_{n,k,m}}
&=& \begin{cases} \mu & \textrm{if}~ I_{n,k,m} = 0 \\
\begin{array}{@{}l}
\mu - a_{k,m} b_{k,m} r_{k,m} \E\Big\{ U'_{n,k,m}\big((1-a_{k,m} e^{-b_{k,m} x_{n,k,m}
\gamma_{n,k}/I_{n,k,m}})
\\[-1mm]
\quad\mbox{}\times
r_{k,m}\big)\gamma_{n,k}e^{-b_{k,m} x_{n,k,m} \gamma_{n,k}/I_{n,k,m}} \Big\} 
\end{array}
&\textrm{otherwise}.
\end{cases}
\label{eq:muCSRA} 
\end{eqnarray}
Clearly, if $I_{n,k,m} = 0$, then $L(\cdot,\cdot,\cdot)$ is an increasing\footnote
  {We use the terms ``increasing'' and ``decreasing'' interchangeably with 
  ``non-decreasing'' and ``non-increasing'', respectively. The terms
  ``strictly-increasing'' and ``strictly-decreasing'' are used when appropriate.
  } function of $x_{n,k,m}$ since $\mu \geq 0$. Therefore, 
$x^*_{n,k,m}(\mu, \vec{I}) = 0$. But if $I_{n,k,m} \neq 0$, then 
$\frac{\partial L(\mu,\vec{I},\vec{x})}{\partial x_{n,k,m}}$ is an 
increasing function of $x_{n,k,m}$ since $U'_{n,k,m}(\cdot)$ is a 
decreasing function of $x_{n,k,m}$. Thus, we have 
\begin{equation}
\mu - a_{k,m} b_{k,m} r_{k,m} \E\Big\{ U'_{n,k,m}\big((1-a_{k,m} e^{-b_{k,m} x_{n,k,m}
\gamma_{n,k}/I_{n,k,m}})r_{k,m}\big)\gamma_{n,k}e^{-b_{k,m} x_{n,k,m} 
\gamma_{n,k}/I_{n,k,m}} \Big\} = 0 
\end{equation} 
for some positive $x_{n, k, m}$ if and only if
$0 \leq \mu \leq a_{k,m} b_{k,m} r_{k,m} U'_{n,k,m}\big((1-a_{k,m})r_{k,m}\big) \E\{\gamma_{n,k}\}$.
Therefore, 
\begin{equation}
x_{n,k,m}^*(\mu,\vec{I}) =
{\begin{cases} \tilde{x}_{n,k,m}(\mu,\vec{I}) & \textrm{if}~ 0 \leq \mu 
\leq a_{k,m} b_{k,m} r_{k,m} U'_{n,k,m}\big((1-a_{k,m})r_{k,m}\big) \E\{\gamma_{n,k}\} \\
0 & \textrm{otherwise}, \end{cases}}
\label{eq:x_n^*}
\end{equation}
where $\tilde{x}_{n,k,m}(\mu,\vec{I})$ satisfies
\begin{eqnarray}
\mu &=& a_{k,m} b_{k,m} r_{k,m} \E\big\{ U'_{n,k,m}\big((1-a_{k,m} 
e^{-b_{k,m} \tilde{x}_{n,k,m}(\mu,\vec{I})
\gamma_{n,k}/I_{n,k,m}})r_{k,m}\big) \gamma_{n,k}e^{-b_{k,m} 
\tilde{x}_{n,k,m}(\mu,\vec{I}) \gamma_{n,k}/I_{n,k,m}} \big\}. 	\nonumber\\[-4mm] \label{eq:xntilde}
\end{eqnarray}
From \eqref{xntilde}, we observe that $\tilde{x}_{n,k,m}(\mu, \vec{I}) 
= \tilde{p}_{n,k,m}(\mu)I_{n,k,m}$, where $\tilde{p}_{n,k,m}(\mu)$ satisfies
\begin{eqnarray}
\mu &=& a_{k,m} b_{k,m} r_{k,m} \E\big\{ U'_{n,k,m}\big((1-a_{k,m} 
e^{-b_{k,m} \tilde{p}_{n,k,m}(\mu)
\gamma_{n,k}})r_{k,m}\big) \gamma_{n,k}e^{-b_{k,m} 
\tilde{p}_{n,k,m}(\mu) \gamma_{n,k}} \big\}. \label{eq:gntilde}
\end{eqnarray}

\begin{figure*}[t]
\renewcommand{\baselinestretch}{1.0}
\newcommand{\pwid}{3in}
\newcommand{\fwid}{3.2in}
\newcommand{\hsp}{2mm}
  \begin{minipage}{\pwid}
  \hspace{\hsp}
    \psfrag{PCSI}[][][0.7]{\sf ~Partial CSI}
	 \psfrag{usermsg}[][][0.7]{$\begin{array}{c} \sf user \\ \sf messages \\ 
	\end{array}$}
	 \psfrag{S&RA}[][][0.7]{$\begin{array}{c} \sf scheduler \\ \sf and \\
	\sf resource \\ \sf allocator \\ 
	\end{array}$}
	\psfrag{n=1}[][][0.7]{\sf $n=1$}
	\psfrag{n=N}[][][0.7]{\sf $n=N$}
	\psfrag{vd}[][][0.7]{\sf $\vdots$}
	\psfrag{vd1}[][][0.7]{$\vdots$}
	\psfrag{MS1}[][][0.7]{\sf ~~user $1$}
	\psfrag{MS2}[][][0.7]{\sf ~~user $2$}
	\psfrag{MSK}[][][0.7]{\sf ~~user $K$}
	\psfrag{BS}[][][0.7]{\sf Base Station}
    \includegraphics[width=\fwid]{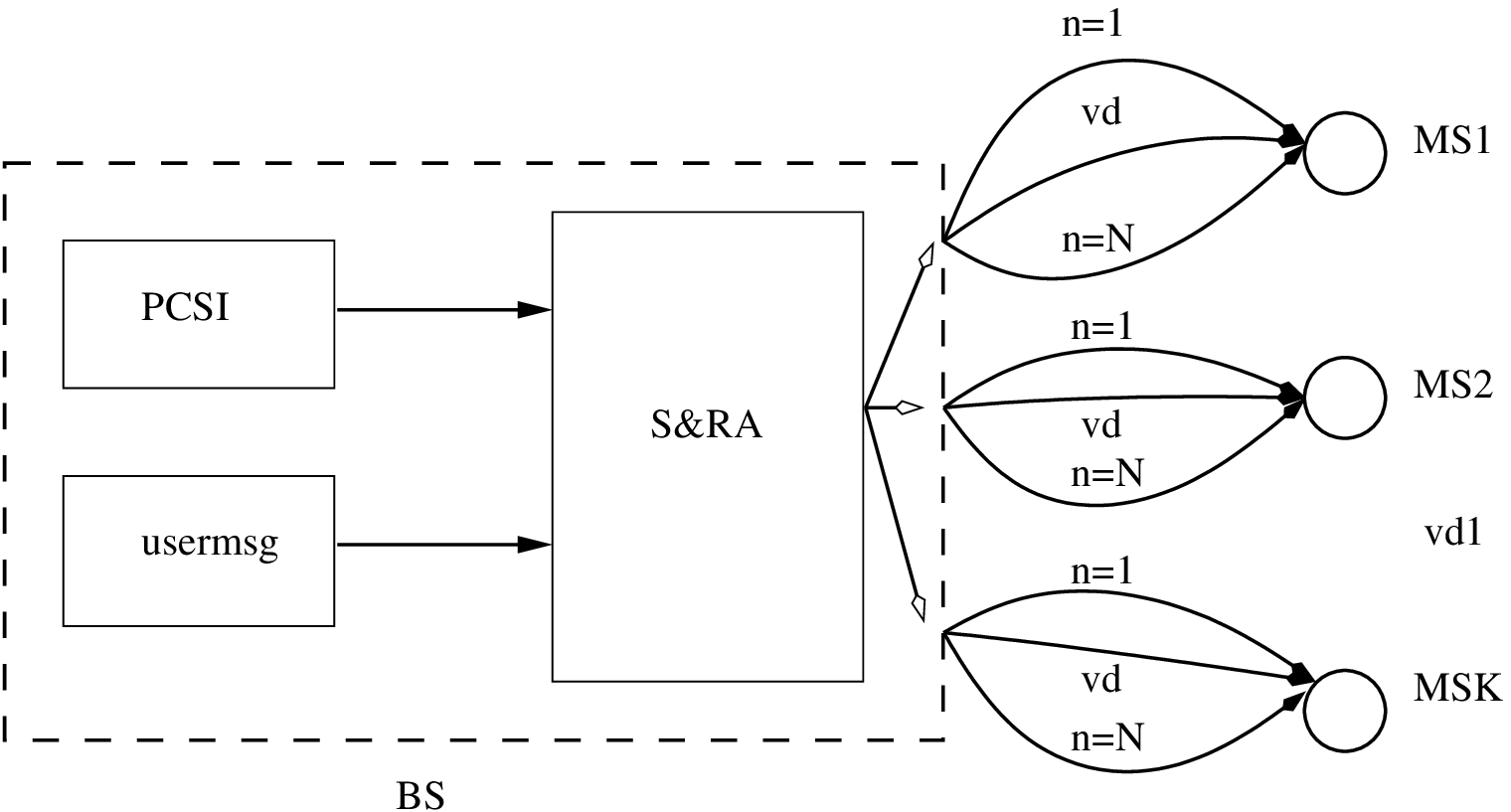}
    \caption[hang]{\footnotesize System model of a downlink OFDMA system with $N$ subchannels 
	and $K$	users. Here, $n$ is the subchannel index.}
    \label{fig:system-model}
  \end{minipage}
  \hspace{11mm}  \hspace{\hsp}  \begin{minipage}{\pwid}
    \psfrag{mu}[t][][0.8]{\sf $\mu$}
	\psfrag{optX}[][][0.7]{$p_{n,k,m}^*(\mu)$}
    \includegraphics[width=\fwid]{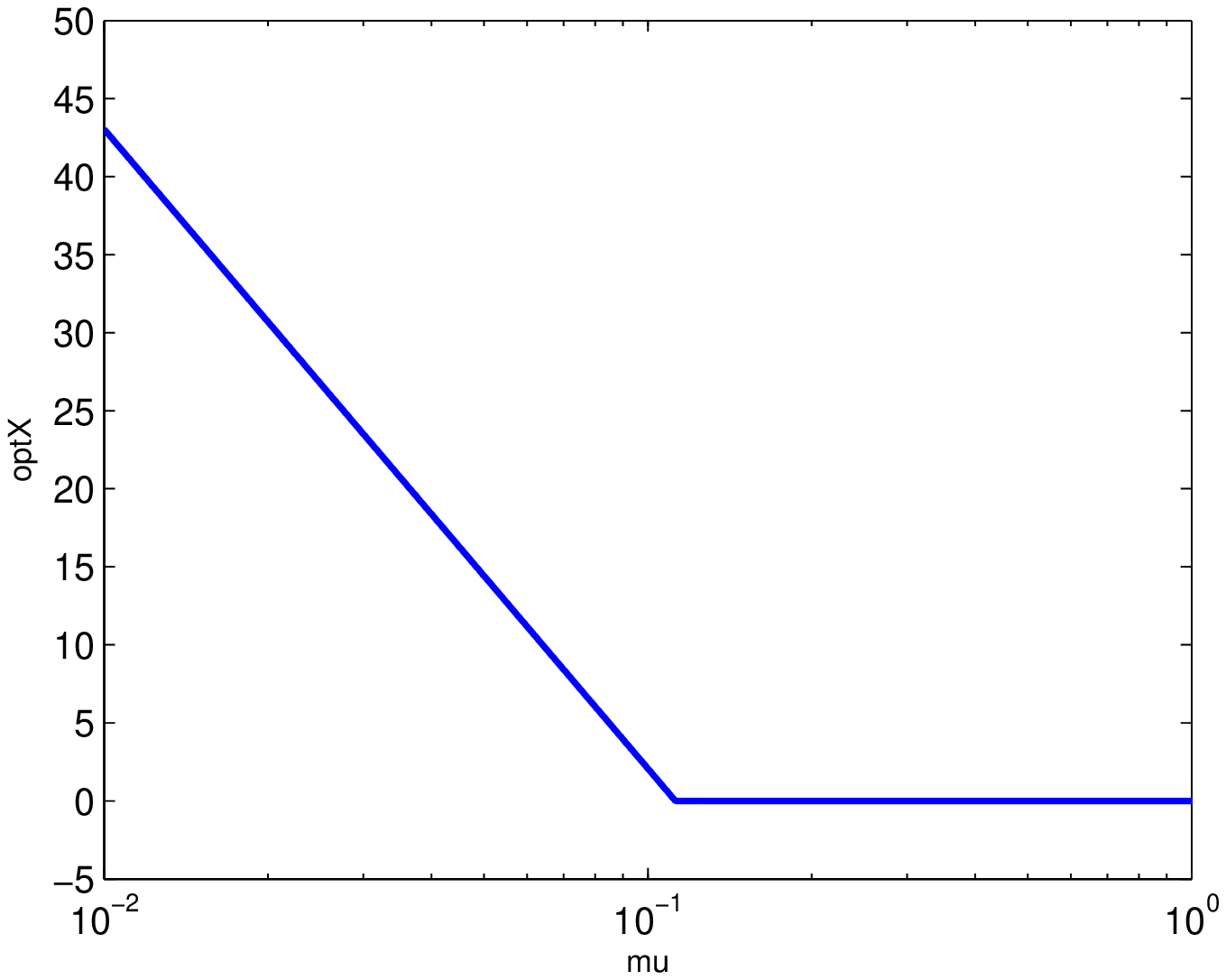}
    \caption[hang]{\footnotesize Prototypical plot of $p_{n,k,m}^*(\mu)$ as a function of $\mu$. 
	The choice of system parameters 
	are the same as those used in \secref{simulations}.}
    \label{fig:optp_versus_mu}
  \end{minipage}
\end{figure*}

Combining the above observations, we can write for any 
$\vec{I} \in \mc{I}\csra$ and $(n,k,m)$ that
\begin{equation}
x_{n,k,m}^*(\mu,\vec{I}) = I_{n,k,m} \, p_{n,k,m}^*(\mu), \label{eq:g_n^*}
\vspace{-4mm}
\end{equation}
where 
\begin{equation}
p_{n,k,m}^*(\mu) =
{\begin{cases} \tilde{p}_{n,k,m}(\mu) & \textrm{if}~ 0 \leq \mu 
\leq a_{k,m} b_{k,m} r_{k,m} U'_{n,k,m}\big((1-a_{k,m})r_{k,m}\big) \E\{\gamma_{n,k}\} \\
0 & \textrm{otherwise},
\end{cases}}
 \label{eq:g*}
\end{equation}
and $\tilde{p}_{n,k,m}(\mu)$ satisfies \eqref{gntilde}.
Note that if such a $\tilde{p}_{n,k,m}(\mu)$ exists that satisfies 
\eqref{gntilde}, then it is unique. This is because, in \eqref{gntilde}, 
$U'_{n,k,m}(\cdot)$ is a continuous decreasing positive function and 
$e^{-b_{k,m} \tilde{p}_{n,k,m}(\mu) \gamma_{n,k}}$ is a strictly-decreasing 
continuous function of $\tilde{p}_{n,k,m}(\mu)$, which makes the right side 
of \eqref{gntilde} a strictly-decreasing continuous function of 
$\tilde{p}_{n,k,m}(\mu)$. Therefore, in the domain of its existence, 
$\tilde{p}_{n,k,m}(\mu)$ is unique and decreases continuously with 
increase in $\mu$. Consequently, $x_{n,k,m}^*(\mu,\vec{I})$ is a decreasing 
continuous function of $\mu$. 
\Figref{optp_versus_mu} shows an example of the variation of 
$p_{n,k,m}^*(\mu)$ w.r.t.\ $\mu$.

\subsection{Optimizing over user-MCS allocation matrix $\vec{I}$ for a given $\mu$}
\label{sec:continuous I}

Substituting $\vec{x}^*(\mu, \vec{I})$ from \eqref{g_n^*} into \eqref{clagrange}, we get the Lagrangian 
\begin{eqnarray}
\lefteqn{L(\mu,\vec{I},\vec{x}^*(\mu,\vec{I}))} \nonumber \\[-6mm]
&=& - \mu P\con + \sum_n \underbrace{\sum_{k,m}
I_{n,k,m} \bigg[ \overbrace{- \E\Big\{ U_{n,k,m}\big((1 - 
a_{k,m} e^{-b_{k,m} p_{n,k,m}^*(\mu) \gamma_{n,k}})r_{k,m} \big)
\Big\} + \mu p_{n,k,m}^*(\mu)}^{V_{n,k,m}(\mu,p_{n,k,m}^*(\mu))} 
\bigg]}_{L_n(\mu,\vec{I}_n)}, \label{eq:OPT1}
\end{eqnarray}
where $\vec{I}_n = \{I_{n,k,m}~\forall (k,m)\}$.
Since the above Lagrangian contains the sum of $L_n(\mu, \vec{I}_n)$ over $n$,
minimizing $L_n(\mu,\vec{I}_n)$ for every $n$ (over all possible 
$\vec{I}_n$) minimizes the Lagrangian.
Recall that $L_n(\mu,\vec{I}_n)$ is a linear function of 
$\{I_{n,k,m}~\forall (k,m)\}$ that satisfies $\sum_{k,m} I_{n,k,m} 
\leq 1$. Therefore, $L_n(\mu,\vec{I}_n)$ is minimized by the
$\vec{I}_n$ that gives maximum possible weight to the 
$(k,m)$ combination with the most negative value 
of $V_{n,k,m}(\mu,p_{n,k,m}^*(\mu))$. To write this 
mathematically, let us define, for each $\mu$ and subchannel 
$n$, a set of participating user-MCS combinations that yield the 
same most-negative value of 
$V_{n,k,m}(\mu,p_{n,k,m}^*(\mu))$ over all $(k,m)$ as follows:
\begin{eqnarray}
S_n(\mu) \defn \Big\{(k,m): (k,m) = \argmin_{(k',m')} V_{n,k',m'}(\mu, 
p_{n,k',m'}^*(\mu)),~\textrm{and}~ V_{n,k,m}(\mu, p_{n,k,m}^*(\mu)) \leq 0\Big\}. \label{eq:Snmu}
\end{eqnarray}
If $S_n(\mu)$ is a null or a singleton set, then
the optimal allocation on subchannel $n$ is given by
\begin{eqnarray}
I_{n,k,m}^*(\mu) = {\begin{cases} 1 & \text{if}~ (k,m) \in S_n(\mu) \\
0 & \text{otherwise}. \end{cases}} \label{eq:I*CSRA}
\end{eqnarray}
However, if $|S_n(\mu)| > 1$ (where $|S_n(\mu)|$ denotes the cardinality of 
$S_n(\mu)$), then multiple $(k,m)$ combinations contribute equally towards the 
minimum value of $L_n(\mu,\vec{I})$, and thus the optimum can be reached by 
sharing subchannel $n$. In particular, let us suppose that 
$S_n(\mu) = \{\big(k_1(n),m_1(n)), \ldots, (k_{|S_n(\mu)|}(n),m_{|S_n(\mu)|}(n)\big)\}$. 
Then, the optimal allocation of subchannel $n$ is given by
\begin{eqnarray}
I_{n,k,m}^*(\mu) = {\begin{cases} I_{n,k_i(n),m_i(n)} & \text{if}~ (k,m) = (k_i(n),m_i(n)) 
~\textrm{for some}~ i \in \{1, \ldots, |S_n(\mu)|\} \\
0 & \text{otherwise}, \end{cases}} \label{eq:I*odd}
\end{eqnarray}
where the vector $(I_{n,k_1(n),m_1(n)}, \ldots, I_{n,k_{|S_n(\mu)|}(n),m_{|S_n(\mu)|}(n)})$ 
is any point in the unit-$(|S_n(\mu)|-1)$ simplex, i.e., it belongs to the space 
$[0,1]^{|S_n(\mu)|}$ and satisfies
\begin{equation}
\sum_{i=1}^{|S_n(\mu)|} I_{n,k_i(n),m_i(n)} = 1. \label{eq:sumI}
\end{equation}

\subsection{Optimizing over $\mu$}
\label{sec:continuous mu}

In order to optimize over $\mu$, we can calculate the Lagrangian optimized 
for a given value of $\mu$ as 
\begin{eqnarray}
\lefteqn{L(\mu,\vec{I}^*(\mu),\vec{x}^*(\mu,\vec{I}^*(\mu)))} \nonumber \\
&=& - \mu P\con + \sum_{n,k,m}
I_{n,k,m}^*(\mu) \bigg[ - \E\Big\{ U_{n,k,m}\big((1 - 
a_{k,m} e^{-b_{k,m} p_{n,k,m}^*(\mu) \gamma_{n,k}})r_{k,m} \big)
\Big\} + \mu p_{n,k,m}^*(\mu) \bigg], \label{eq:optI_mu}
\end{eqnarray}
and then maximize it over all possible values of $\mu \geq 0$ to find 
$\mu^*$. Notice from \eqref{I*CSRA}-\eqref{sumI} that we have 
$\sum_{k,m}I_{n,k,m}^*(\mu^*) = 1$ for at least one $n$. Otherwise, $\vec{I}^*(\mu^*) = \vec{0}$
which, clearly, is not the optimal solution. 
Therefore, $\mu^* \geq \mu_{\text{\sf min}} > 0$, where
\begin{equation}
\mu_{\text{\sf min}} = \min_{n,k,m} 
a_{k,m} b_{k,m} r_{k,m} \E\big\{U_{n,k,m}'\big((1-a_{k,m} e^{-b_{k,m} P\con 
\gamma_{n,k}})r_{k,m}\big)\gamma_{n,k}e^{-b_{k,m} P\con \gamma_{n,k}}\big\} \label{eq:mumin}
\end{equation}
is obtained by taking $\tilde{p}_{n,k,m}(\mu)\rightarrow P\con$ 
for all $(n,k,m)$ in the right side of \eqref{gntilde}. 
Since $p^*_{n,k,m}(\mu)$ is a decreasing continuous function of $\mu$ 
(seen in \secref{continuous p}), we have 
$\sum_{n,k,m}x_{n,k,m}^*(\mu,\vec{I}) > P\con$ for all $\vec{I} \neq \vec{0}$ 
and $\mu < \mu_{\text{\sf min}}$.
We can also obtain an upper bound 
$\mu^* \leq \mu_{\text{\sf max}}$, where 
\begin{equation}
\mu_{\text{\sf max}} 
= \max_{n,k,m} a_{k,m} b_{k,m} r_{k,m} U_{n,k,m}'\big((1-a_{k,m})r_{k,m}\big) 
\E\{\gamma_{n,k}\} \label{eq:mumax}
\end{equation}
is obtained by taking $\tilde{p}_{n,k,m}(\mu)\rightarrow 0$ in the right 
side of \eqref{gntilde}. Thus, for any 
$\mu > \mu_{\text{\sf max}}$, we have that
$x_{n,k,m}^*(\mu,\vec{I}) = 0 ~\forall n, k, m, \vec{I}$. 
Since the primal objective
in \eqref{CSRAP} is certainly not maximized when zero power is allocated on 
all subchannels, we have $\mu^* \in [\mu_{\text{\sf min}}, 
\mu_{\text{\sf max}}] \subset (0, \infty)$. 

At the optimal $\mu$, i.e., $\mu^*$, if we have $|S_n(\mu^*)| 
\leq 1~\forall n$, then the optimal CSRA allocation, $\vec{I}^*\csra$, 
equals $\vec{I}^*(\mu^*)$ and can be calculated using 
\eqref{I*CSRA}. Moreover, the optimal power allocation $\vec{p}^*\csra$ allocates 
\begin{equation}
p_{n,k,m, \text{\sf CSRA}}^* = 
{\begin{cases} 
p_{n,k,m}^*(\mu^*) & \textrm{if}~ 
I^*_{n,k,m}(\mu^*) \neq 0 \\ 
0 & \textrm{otherwise} \end{cases}} \label{eq:optP}
\end{equation}
to every possible $(n,k,m)$ combination. However, if for some $n$, 
we have $|S_n(\mu^*)| > 1$, then ambiguity arises due to multiple possibilities of
$\vec{I}^*(\mu^*)$ obtained via \eqref{I*odd}. In order to find the 
optimal user-MCS allocation in such cases, we use the fact that the CSRA 
problem in \eqref{CSRAP} is a convex optimization problem whose exact 
solution satisfies the sum-power constraint with equality, i.e.,
\begin{equation}
\sum_{n,k,m}x_{n,k,m}^*(\mu^*,\vec{I}^*(\mu^*)) = 
\sum_{n,k,m}I_{n,k,m}^*(\mu^*)p_{n,k,m}^*(\mu^*) = P\con.
\end{equation}
This is because $\mu^* \geq \mu_{\text{\sf min}} > 0$ (shown earlier) and the complementary 
slackness condition gives that 
$\mu^* \big(\sum_{n,k,m}x_{n,k,m}^*(\mu^*,\vec{I}^*(\mu^*)) - P\con\big) = 0$.
Now, recall that the total 
power allocated to any subchannel 
$n$ at $\mu^*$ is $\sum_{i=1}^{|S_n(\mu^*)|} I_{n,k_i(n),m_i(n)} \, 
p_{n,k_i(n),m_i(n)}^*(\mu^*)$ where $\{I_{n,k_i(n),m_i(n)}\}_{i=1}^{|S_n(\mu^*)|}$ 
satisfies \eqref{sumI}. This quantity is dependent on the choice of values for 
$\{I_{n,k_i(n),m_i(n)}\}_{i=1}^{|S_n(\mu^*)|}$ and takes on any value between 
an upper and lower bound given by the following equation:
\begin{equation}
\min_{i} \, p_{n,k_i(n),m_i(n)}^*(\mu^*) \leq \sum_{i=1}^{|S_n(\mu^*)|} 
I_{n,k_i(n),m_i(n)} \, p_{n,k_i(n),m_i(n)}^*(\mu^*) \leq 
\max_{i} \, p_{n,k_i(n),m_i(n)}^*(\mu^*). \label{eq:key}
\end{equation}
Note that the existence of at least one $\vec{I} = \vec{I}^*(\mu^*)$ satisfying 
\begin{equation}
\sum_n \sum_i I_{n,k_i(n),m_i(n)} \, p_{n,k_i(n),m_i(n)}^*(\mu^*) = P\con
\label{eq:final}
\end{equation}
is guaranteed by the optimality of the dual solution (of our convex CSRA problem 
over a closed constraint set). Therefore, we necessarily have
$\sum_n \min_{i}p_{n,k_i(n),m_i(n)}^*(\mu^*) \leq P\con$, and
$\sum_n \max_{i}p_{n,k_i(n),m_i(n)}^*(\mu^*) \geq P\con$.
In addition, all choices of user-MCS allocations, $\vec{I}^*(\mu^*)$, given by 
\eqref{I*odd} that satisfy the equality 
$\sum_{n,k,m} I^*_{n,k,m}(\mu^*) \, p_{n,k,m}^*(\mu^*) = P\con$,
are optimal for the CSRA problem. 

In the case that the optimal solution $\vec{I}^*(\mu^*)$ is non-unique, i.e., 
$|S_n(\mu^*)| > 1$ for some $n$, then one instance of $\vec{I}^*(\mu^*)$ can be
found as follows. For each subchannel $n$, define
\begin{eqnarray}
(k_{\scriptscriptstyle\text{\sf max}}(n,\mu^*), m_{\scriptscriptstyle\text{\sf max}}(n,\mu^*)) 
&:=& \argmax_{i}p_{n,k_i(n),m_i(n)}^*(\mu^*), \\
(k_{\scriptscriptstyle\text{\sf min}}(n,\mu^*), m_{\scriptscriptstyle\text{\sf min}}(n,\mu^*)) 
&:=& \argmin_{i}p_{n,k_i(n),m_i(n)}^*(\mu^*),
\end{eqnarray}
and find the value of $\lambda \in [0, 1]$ for which
\begin{equation}
\lambda \Big(\sum_n p_{n, k_{\scriptscriptstyle\text{\sf min}}(n,\mu^*),
m_{\scriptscriptstyle\text{\sf min}}(n,\mu^*)}(\mu^*)\Big) + 
(1-\lambda)\Big(\sum_n p_{n, k_{\scriptscriptstyle\text{\sf max}}(n,\mu^*),
m_{\scriptscriptstyle\text{\sf max}}(n,\mu^*)}(\mu^*)\Big) = P\con,
\vspace{-3mm}
\end{equation}
\ie
\begin{equation}
\lambda = \frac{\sum_n p_{n, k_{\scriptscriptstyle\text{\sf max}}(n,\mu^*), 
m_{\scriptscriptstyle\text{\sf max}}(n,\mu^*)}(\mu^*)-P\con}
{\sum_n p_{n, k_{\scriptscriptstyle\text{\sf max}}(n,\mu^*), 
m_{\scriptscriptstyle\text{\sf max}}(n,\mu^*)}(\mu^*) - \sum_n 
p_{n, k_{\scriptscriptstyle\text{\sf min}}(n,\mu^*), 
m_{\scriptscriptstyle\text{\sf min}}(n,\mu^*)}(\mu^*)}.
\end{equation}
Now, defining two specific allocations, $\vec{I}^{\scriptscriptstyle\text{\sf min}}(\mu^*)$
and $\vec{I}^{\scriptscriptstyle\text{\sf max}}(\mu^*)$, as
\begin{eqnarray}
I_{n,k,m}^{\scriptscriptstyle\text{\sf min}}(\mu^*) = \small{\begin{cases} 1 &  (k,m) = (k_{\scriptscriptstyle\text{\sf 
min}}(n,\mu^*), m_{\scriptscriptstyle\text{\sf min}}(n,\mu^*)) \\
0 & \text{otherwise}, \end{cases}}\!\!,\,\,
I_{n,k,m}^{\scriptscriptstyle\text{\sf max}}(\mu^*) = \small{\begin{cases} 1 & (k,m) = (k_{\scriptscriptstyle\text{\sf max}}(n,\mu^*), 
m_{\scriptscriptstyle\text{\sf max}}(n,\mu^*)) \\
0 & \text{otherwise}, \end{cases}} 
\nonumber\\[-7mm]\label{eq:Imaximin}
\end{eqnarray}
respectively, the optimal user-MCS allocation is given by 
$\vec{I}^*\csra = \lambda \vec{I}^{\scriptscriptstyle\text{\sf min}}(\mu^*) + 
(1-\lambda)\vec{I}^{\scriptscriptstyle\text{\sf max}}(\mu^*)$.
The corresponding optimal power allocation is then given by \eqref{optP}. 
It can be seen that this solution satisfies the subchannel constraint 
as well as the sum power constraint with equality, i.e.,
\begin{equation}
\sum_{n,k,m}I_{n,k,m}^*(\mu^*)p_{n,k,m}^*(\mu^*) = 
\sum_{n,k,m}x_{n,k,m}^*(\mu^*, \vec{I}^*(\mu^*)) = P\con. \nonumber
\end{equation}

Two interesting observations can be made from the above discussion. Firstly, 
for any choice of concave utility functions $U_{n,k,m}(\cdot)$, there exists an 
optimal scheduling and resource allocation strategy that allocates each subchannel
to at most $2$ user-MCS combinations. Therefore, when allocating $N$ subchannels, 
even if more than $2N$ user-MCS options are available, at most $2N$ 
such options will be used.
Secondly, if $\vec{I}^{\scriptscriptstyle\text{\sf min}}(\mu^*) = 
\vec{I}^{\scriptscriptstyle\text{\sf max}}(\mu^*)$, then the exact CSRA solution allocates 
power to at most one $(k,m)$ combination for every subchannel, i.e., no subchannel 
is shared among any two or more user-MCS combinations. This 
observation will motivate the SRA problem's solution without subchannel 
sharing in \secref{discrete}.

\subsection{Algorithmic implementation} \label{sec:continuous imple}

In practice, it is not possible to search exhaustively over
$\mu \in [\mu_{\text{\sf min}}, 
\mu_{\text{\sf max}}]$. 
Thus, we propose an algorithm to reach solutions in close (and adjustable) 
proximity to the optimal. 
The algorithm first narrows down the location of $\mu^*$ 
using a bisection-search over $[\mu_{\text{\sf min}}, 
\mu_{\text{\sf max}}]$ for the optimum total power 
allocation, and then finds a set of resource allocations
$(\vec{I}, \vec{x})$ that achieve a total utility close to the optimal. 

To proceed in this direction, with the aim of developing a framework to do
bisection-search over $\mu$, let us define the total optimal allocated 
power for a given value of $\mu$ as follows:
\begin{eqnarray}
X^*\overall(\mu)
&\defn& \sum_{n,k,m} x^*_{n,k,m}(\mu,\vec{I}^*(\mu)),
\label{eq:xoverall}
\end{eqnarray}
where $\vec{I}^*(\mu)$ and $\vec{x}^*(\mu, \vec{I}^*(\mu))$ (defined in 
\eqref{cdual}) minimize the Lagrangian (defined in \eqref{clagrange}) for 
a given $\mu$. The following lemma relates the variation of $X^*\overall(\mu)$ 
with respect to $\mu$.
\begin{lemma} \label{lem:lem3}
The total optimal power allocation, $X\overall^*(\mu)$, is a
monotonically decreasing function of $\mu$.
\end{lemma}
\begin{IEEEproof}
A proof sketch is given in Appendix \ref{appendix2}. 
For the full proof, see \cite{mythesis}.
\end{IEEEproof}
A sample plot of $X^*\overall(\mu)$ and $L(\mu, \vec{I}^*(\mu),
\vec{x}^*(\mu, \vec{I}^*(\mu)))$ as a function of $\mu$ is shown in \Figref{sump}.
From the figure, three observations can be made. First, as $\mu$ increases, 
the optimal total allocated power decreases, as expected from 
\lemref{lem3}. Second, as expected, the Lagrangian is maximized for that 
value of $\mu$ at which $X^*\overall(\mu) = P\con$. Third, the optimal total
power allocation varies continuously in the region of $\mu$ where the 
optimal allocation, $\vec{I}^*(\mu)$, remains constant and takes a jump 
(negative) when $\vec{I}^*(\mu)$ changes. This happens for the 
following reason. We know, for any $(n,k,m)$, that $p_{n,k,m}^*(\mu)$ is a 
continuous function of $\mu$. Thus, when the optimal allocation remains constant
over a range of $\mu$, the total power allocated, 
$\sum_{n,k,m}I_{n,k,m}^*(\mu)p_{n,k,m}^*(\mu)$ also varies continuously 
with $\mu$. However, at the point of discontinuity (say $\tilde{\mu}$), 
multiple optimal allocations achieve the same optimal value of Lagrangian. 
In other words, $|S_n(\tilde{\mu})| > 1$ for some $n$. In that case, 
$X^*\overall(\tilde{\mu})$ can take any value in the interval
\begin{equation}\textstyle
\Big[ \sum_{n} p_{n,k_{\scriptscriptstyle\text{\sf min}}(n), 
m_{\scriptscriptstyle\text{\sf min}}(n)}^*(\tilde{\mu}),~
\sum_{n} p_{n,k_{\scriptscriptstyle\text{\sf max}}(n), 
m_{\scriptscriptstyle\text{\sf max}}(n)}^*(\tilde{\mu})\Big] \nonumber
\end{equation}
while achieving the same minimum value of the Lagrangian at 
$\tilde{\mu}$. Applying \lemref{lem3}, we have
\begin{eqnarray}
X^*\overall(\tilde{\mu}-\Delta_1) 
~\geq~ 
\sum_{n} p_{n,k_{\scriptscriptstyle\text{\sf max}}(n), 
m_{\scriptscriptstyle\text{\sf max}}(n)}^*(\tilde{\mu}) 
~\geq~
X^*\overall(\tilde{\mu})
~\geq~ 
\sum_{n} p_{n,k_{\scriptscriptstyle\text{\sf min}}(n), 
m_{\scriptscriptstyle\text{\sf min}}(n)}^*(\tilde{\mu}) 
~\geq~
X^*\overall(\tilde{\mu}+\Delta_2) \nonumber
\end{eqnarray}
for any $\Delta_1, \Delta_2 > 0$, causing a jump of
$\big(\sum_{n} p_{n,k_{\scriptscriptstyle\text{\sf min}}(n), 
m_{\scriptscriptstyle\text{\sf min}}(n)}^*(\tilde{\mu})
- \sum_{n} p_{n,k_{\scriptscriptstyle\text{\sf max}}(n), 
m_{\scriptscriptstyle\text{\sf max}}(n)}^*(\tilde{\mu})\big)$
in the total optimal power allocation at $\tilde{\mu}$.

\newcommand{\colwidth}{8.0cm}
\begin{table}[t]
\renewcommand{\baselinestretch}{1.0}

\caption{Algorithmic implementations of the proposed algorithms}
\begin{tabular}{|@{}p{\colwidth}|@{}p{\colwidth}|}
  \hline
\small{~~Proposed CSRA algorithm} & 
\small{~~Brute force algorithm for a given $\vec{I}$} \\ \hline
\begin{enumerate}\footnotesize
  \item Set $\underline{\mu} = \mu_{\text{\sf min}}$, 
  	$\bar{\mu} = \mu_{\text{\sf max}}$, and 
  	$\mu = \frac{\underline{\mu}+\bar{\mu}}{2}$.
   \item For each subchannel $n=1,\dots,N$:
   \begin{enumerate}
    \item For each $(k,m)$,
    \begin{enumerate}
     \item Use \eqref{gntilde} and \eqref{g*} to calculate 
    	$p_{n,k,m}^*(\mu)$.
     \item Use \eqref{OPT1} to calculate 
	$V_{n,k,m}(\mu, p_{n,k,m}^*(\mu))$.
    \end{enumerate}
   \item Calculate $S_n(\mu)$ using \eqref{Snmu}. 
   \end{enumerate}
   \item If $|S_n(\mu)| \leq 1~\forall n$, then find $\vec{I}^*(\mu)$ using \eqref{I*CSRA},
        else use \eqref{Imaximin} and set 
	$\vec{I}^*(\mu) = \vec{I}^{\scriptscriptstyle\text{\sf min}}(\mu)$. 
   \item Find $\vec{x}^*(\mu, \vec{I}^*(\mu))$ 
	using \eqref{g_n^*} and calculate 
$X\overall^*(\mu) = \sum_{n,k,m} x^*_{n,k,m}(\mu, \vec{I}^*(\mu))$.
   \item If $X\overall^*(\mu) \geq P\con$, 
	set $\underline{\mu} = \mu$, 
   otherwise set $\bar{\mu} = \mu$.
   \item If $\bar{\mu} - \underline{\mu} > \kappa$, go to 
   step 2), else proceed.
 \item Now we have $\mu^* \in [\underline{\mu}, 
 \bar{\mu}]$ and $\bar{\mu} - \underline{\mu} < \kappa$. If 
$X\overall^*(\underline{\mu}) \neq X\overall^*(\bar{\mu})$, set
 $\lambda = \frac{X\overall^*(\underline{\mu}) - P\con }{X\overall^*(\underline{\mu}) - 
X\overall^*(\bar{\mu})}$, else set $\lambda = 0$.
 \item The optimal user-MCS allocation is given by 
	$\vec{\hat{I}}_{\text{\sf CSRA}} = 
	\lambda \vec{I}^*(\bar{\mu})+ (1-\lambda) \vec{I}^*(\underline{\mu})$ 
	and the corresponding optimal $\vec{x}$
	is given by $\vec{\hat{x}}_{\text{\sf CSRA}}
	= \lambda \vec{x}^*(\bar{\mu},\vec{I}^*(\bar{\mu})) + 
	(1-\lambda) \vec{x}^*(\underline{\mu},\vec{I}^*(\underline{\mu}))$. 
	The optimal power allocation, 
	$\vec{\hat{p}}_{\text{\sf CSRA}}$, then can be found using
	\begin{eqnarray}
	\scriptstyle
	\hat{p}_{n,k,m, \text{\sf CSRA}} =
	\small{\left\{ \begin{array}{ll}
	\frac{\hat{x}_{n,k,m, \text{\sf CSRA}}}
	{\hat{I}_{n,k,m, \text{\sf CSRA}}} & \textrm{if}~ 
	\hat{I}_{n,k,m, \text{\sf CSRA}} \neq 0 \\
	0 & \textrm{otherwise}, \end{array}\right.}
	\end{eqnarray}
where $\hat{I}_{n,k,m, \text{\sf CSRA}}$ and $\hat{x}_{n,k,m, \text{\sf CSRA}}$ 
denote the $(n,k,m)^{\textrm{th}}$ component of $\vec{\hat{I}}_{\text{\sf CSRA}}$ 
and $\vec{\hat{x}}_{\text{\sf CSRA}}$, respectively. Notice 
that the obtained solution satisfies the sum-power constraint with equality.
\end{enumerate}
& 
\begin{enumerate}\footnotesize
\item Initialize $\underline{\mu} = \mu_{\text{\sf min}}$ and
$\bar{\mu} = \mu_{\text{\sf max}}$.
\item Set $\mu = \frac{\underline{\mu} + \bar{\mu}}{2}$.
\item For each $(n,k,m)$,
	use \eqref{pn*brute}-\eqref{mubrute} to obtain $x_{n,k,m}^*(\mu)$.
\item Find $X\overall^*(\vec{I},\mu)$ using \eqref{Xtotal}. 
\item If $X\overall^*(\vec{I},\mu) > P\con\,$, 
set $\underline{\mu} = \mu$, otherwise set $\bar{\mu} = \mu$.
\item If $\bar{\mu} - \underline{\mu} < \kappa$, go to step 7), otherwise go to step 2).
\item If $X\overall^*(\vec{I},\bar{\mu}) \neq X\overall^*(\vec{I},\underline{\mu})$, set 
$\lambda = \frac{X\overall^*(\vec{I},\underline{\mu}) - P\con}{X\overall^*(\vec{I},\underline{\mu}) - 
X\overall^*(\vec{I},\bar{\mu})}$, otherwise set $\lambda = 0$. 
\item Set $\hat{\mu}_{\vec{I}} = \bar{\mu}$. The best actual power allocation is given by 
$\vec{\hat{x}}_{\vec{I}} = \lambda \vec{x}^*(\bar{\mu})
+ (1-\lambda) \vec{x}^*(\underline{\mu})$ and the best power allocation, $\vec{\hat{p}}_{\vec{I}}$, is given by
\begin{equation}
\hat{p}_{n,k,m, \vec{I}} = \small{\begin{cases} \frac{\hat{x}_{n,k,m, \vec{I}}}{I_{n,k,m}} & 
\textrm{if}~I_{n,k,m}\neq 0 \\
0 & \textrm{otherwise}, \end{cases}} \nonumber
\end{equation}
where $\hat{p}_{n,k,m, \vec{I}}$ and $\hat{x}_{n,k,m, \vec{I}}$ are the $(n,k,m)^{\textrm{th}}$ element
of $\vec{\hat{p}}_{\vec{I}}$ and $\vec{\hat{x}}_{\vec{I}}$, respectively.
The corresponding Lagrangian, found using $\hat{L}_{\vec{I}} = L_{\vec{I}}(\bar{\mu}, \vec{p}^*(\mu))$,
gives the optimal Lagrangian value.
\end{enumerate}
\vspace{3mm}
\begin{tabular}{@{}p{\colwidth}} 
\hline 
\small{~~Proposed DSRA algorithm}\\
\hline
\begin{enumerate}\footnotesize
  \item Use the algorithmic implementation of the proposed CSRA solution in 
	to find $\vec{I}^*(\underline{\mu})$ 
	and $\vec{I}^*(\bar{\mu})$, where the optimal $\mu$
	for the CSRA problem, i.e., $\mu^*$ lies in the set $[\underline{\mu}, \bar{\mu}]$,
	$\bar{\mu} - \underline{\mu} < \kappa$, and $\vec{I}^*(\underline{\mu}), 
	\vec{I}^*(\bar{\mu}) \in \mc{I}\dsra$.
\item For both $\vec{I} = \vec{I}^*(\underline{\mu})$ and $\vec{I}=\vec{I}^*(\bar{\mu})$ (since they may differ),
  calculate $\vec{\hat{p}}_{\vec{I}}$ and 
	$\hat{L}_{\vec{I}}$ as described for the brute force algorithm.
  \item Choose $\vec{\hat{I}}\dsra 
	= \argmin_{\vec{I} \in \{\vec{I}^*(\underline{\mu}),\, \vec{I}^*(\bar{\mu})\}} \hat{L}_{\vec{I}}$ 
	as the user-MCS allocation 
	and $\vec{\hat{p}}\dsra = \vec{\hat{p}}_{\vec{\hat{I}}\dsra}$ as the associated 
	power allocation.
\end{enumerate} 
\end{tabular}
\\ \hline
\end{tabular}
\label{table1}
\vspace{-5mm}
\end{table}

\lemref{lem3} allows us to do a bisection-search over $\mu$ since $X\overall^*(\mu)$ 
is a decreasing function of $\mu$ and the optimal $\mu$ 
is the one at which $X\overall^*(\mu) = P\con$. In particular, 
if $\mu^* \in [\underline{\mu}, \bar{\mu}]$ for some $\underline{\mu}$ 
and $\bar{\mu}$, then 
$\mu^* \in \left[\frac{\underline{\mu}+\bar{\mu}}{2}, \bar{\mu}\right]$
if $X\overall^*\left(\frac{\underline{\mu}+\bar{\mu}}{2}\right) > P\con$, otherwise
$\mu^* \in \left[\underline{\mu}, \frac{\underline{\mu}+\bar{\mu}}{2}\right]$.
Using this concept, we propose an algorithm in Table \ref{table1} that finds 
an interval $[\underline{\mu}, \bar{\mu}]$, such that $\mu^* \in [\underline{\mu}, 
\bar{\mu}]$ and $\bar{\mu} - \underline{\mu} \leq \kappa$, where $\kappa~(>0)$ is a
tuning-parameter, and allocates resources based on optimal 
resource allocations at $\underline{\mu}$ and $\bar{\mu}$. 

The following lemma characterizes the relationship between the tuning parameter $\kappa$
and the accuracy of the obtained solution.
\begin{lemma} \label{lem:lem6}
Let $\mu^* \in [\underline{\mu}, \bar{\mu}]$ be the point where the proposed 
CSRA algorithm stops, and the total utility obtained by the 
proposed algorithm and the exact CSRA solution be 
$\hat{U}_{\text{\sf CSRA}}(\underline{\mu}, \bar{\mu})$ 
and $U^*\csra$, respectively. Then,
$0 \leq U^*\csra - \hat{U}_{\text{\sf CSRA}}(\underline{\mu}, \bar{\mu}) \leq 
(\bar{\mu}-\underline{\mu})P\con$.
\end{lemma}
\begin{IEEEproof}
For proof, see Appendix \ref{appendix7}.
\end{IEEEproof}
Since our algorithm stops when $\bar{\mu}-\underline{\mu} \leq \kappa$,
from \lemref{lem6}, the gap between the obtained utility and the optimal 
utility is bounded by $P\con \kappa$. Moreover, 
$\lim_{\underline{\mu} \to \bar{\mu}} 
\hat{U}_{\text{\sf CSRA}}(\underline{\mu}, \bar{\mu}) = U^*\csra$. 

The proposed algorithm requires at most $\left\lceil \log_2 
\big(\frac{\mu_{\text{\sf max}} - 
\mu_{\text{\sf min}}}{\kappa}\big) \right\rceil$
iterations of $\mu$ in order to find $\bar{\mu}$, and $\underline{\mu}$ 
such that $\bar{\mu} - \underline{\mu} \leq \kappa$ and $\mu^* \in 
[ \underline{\mu}, \bar{\mu}]$. Therefore, measuring the complexity of the
algorithm by the number of times \eqref{gntilde} must be solved for a given
$(n,k,m,\mu)$, the proposed algorithm takes at most
\begin{equation}\textstyle
NKM \left\lceil \log_2 \big(\frac{\mu_{\text{\sf max}} - 
\mu_{\text{\sf min}}}{\kappa}\big) \right\rceil 
\label{eq:complexitycsra}
\end{equation}
steps. We use this method of measuring complexity because it allows
us to easily compare all algorithms in the paper. 
Note that, for a given $\kappa$, the number of steps taken by the proposed 
bisection algorithm is proportional to $\log_2{\kappa}$.

\subsection{Some properties of the CSRA solution} 
\label{sec:continuous prop}

In this subsection, we study a few properties of the CSRA solution that
yield valuable insights into the optimal resource allocation strategy 
for any given 
value of Lagrange multiplier, $\mu$. Let us fix a $\tilde{\mu} \in 
[\mu_{\text{\sf min}}, \mu_{\text{\sf max}}]$. 
Now, if $|S_n(\tilde{\mu})| \leq 1, \forall n$, then the optimal allocation at 
$\tilde{\mu}$, $\vec{I}^*(\tilde{\mu})$, is given by \eqref{I*CSRA},
which reveals that $\vec{I}^*(\tilde{\mu}) \in \{0, 1\}^{N \times K \times M}$.
In this case, the definition of $\mc{I}\csra$ implies that every subchannel 
is allocated to at most one user-MCS combination. Note that this is precisely the 
constraint we impose in the later part of this paper. Let us now consider the 
case where it is possible that $|S_n(\tilde{\mu})| > 1$ for some $n$.
\begin{lemma} \label{lem:lem1}
For any $\tilde{\mu} > 0$, there exists a $\delta > 0$ such that 
for all $\mu \in (\tilde{\mu}-\delta, \tilde{\mu}+\delta) \setminus 
\{\tilde{\mu}\}$, there exists an optimal allocation,
$\vec{I}^*(\mu) \in \mc{I}\csra$, that satisfies
$\vec{I}^*(\mu) \in \{0, 1\}^{N \times K \times M}$.
Moreover, if $\mu_1,\mu_2 \in (\tilde{\mu} - \delta, \tilde{\mu})$, then there 
exists $\vec{I}^*(\mu_1), \vec{I}^*(\mu_2) \,\in \{0, 1\}^{N \times 
K \times M}$ such that $\vec{I}^*(\mu_1) = \vec{I}^*(\mu_2)$. The same property
holds if both $\mu_1,\mu_2 \in (\tilde{\mu}, \tilde{\mu} + \delta)$. 
\end{lemma}
\begin{IEEEproof}
A proof sketch is given in Appendix \ref{appendix6}. 
For the full proof, see \cite{mythesis}.
\end{IEEEproof}
In conjunction with \eqref{g_n^*}, the above lemma implies that 
the discontinuities in \figref{sump} are isolated and that, around 
every point on the horizontal axis, 
there is a small region over which $X\overall^*(\mu)$ is continuous.
Hence, the number of such discontinuities are, at most, countable.



\section{Scheduling and Resource Allocation without subchannel sharing} 
\label{sec:discrete}

In this section, we will solve the scheduling and resource allocation 
(SRA) problem \eqref{SRA} under the constraint that $I_{n,k,m} 
\in \{0,1\}$, i.e., that each subchannel can be 
allocated to at most one combination of user and MCS per time slot. 
We will refer to this problem as the ``discrete scheduling and resource 
allocation'' (DSRA) problem.
Storing the values of $I_{n,k,m}$ in the $N \times K \times M$ matrix
$\vec{I}$, the DSRA subchannel constraint can be expressed as 
$\vec{I} \in \mathcal{I}\dsra$, where
\begin{equation}
\mathcal{I}\dsra := \bigg\{\vec{I} : \vec{I} \in \{0,1\}^{N \times K \times 
M}, \sum_{k,m}I_{n,k,m} \leq 1 ~\forall n\bigg\}. \nonumber
\end{equation}
Then, using \eqref{SRA}, the DSRA problem can be stated as
\begin{eqnarray}
\text{DSRA} &:=& \max_{\substack{\scriptstyle \{p_{n,k,m} \geq 0\}\\[0.8 mm]
	\scriptstyle \vec{I} \in \mathcal{I}\dsra}} 
	\sum_{n,k,m} \hspace{-1mm}I_{n,k,m}
	\E\Big\{ U_{n,k,m}\big((1-a_{k,m} e^{-b_{k,m} p_{n,k,m} \gamma_{n,k}})r_{k,m}\big) 
	\Big\} 						
~\text{s.t.} \sum_{n,k,m} I_{n,k,m} p_{n,k,m} \leq P\con.
\nonumber\\[-7mm]\label{eq:DSRA}
\end{eqnarray}
Let us denote the 
optimal $\vec{I}$ and $\vec{p}$ for \eqref{DSRA} by $\vec{I}^*\dsra$
and $\vec{p}^*\dsra$, respectively.

The DSRA problem is a mixed-integer programming problem.
Mixed-integer programming problems are generally NP-hard, meaning that 
polynomial-time solutions do not exist~\cite{Lee:Book:01}. Fortunately,
in some cases, such as ours, one can exploit the problem
structure to design polynomial-complexity algorithms that 
reach solutions in close vicinity of the exact solution. We first 
describe an approach to solve the DSRA mixed-integer programming problem 
exactly by exhaustively searching over all possible user-MCS 
allocations in order to arrive at the optimal user, rate,
and power allocation. We will see that this ``brute-force'' approach
has a complexity that grows exponentially in the number
of subchannels. Later, we will exploit the DSRA problem structure,
and its relation to the CSRA problem, to design an algorithm with 
near-optimal performance and polynomial complexity. 

\subsection{Brute-force algorithm}		\label{sec:brute}

Consider that, if we attempted to solve our DSRA problem via 
brute-force (\ie by solving the power allocation sub-problem 
for every possible choice of $\vec{I}\in\mathcal{I}\dsra$), 
we would solve the following sub-problem for every given $\vec{I}$.
\begin{eqnarray}
&& \max_{\scriptstyle \{p_{n,k,m} \geq 0\}} \sum_{n,k,m} 
	 I_{n,k,m}
	\E\Big\{ U_{n,k,m}\big((1-a_{k,m} e^{-b_{k,m} p_{n,k,m} \gamma_{n,k}})r_{k,m}\big) 
	\Big\} 	
~~\text{s.t.} \sum_{n,k,m} I_{n,k,m} \, p_{n,k,m} \leq P\con.\nonumber\\[-5mm]
\label{eq:subsubproblem}
\end{eqnarray}
Borrowing our approach to the CSRA problem, we could transform
the variable $p_{n,k,m}$ into $x_{n,k,m}$ via the relation: $x_{n,k,m} = 
I_{n,k,m}\,p_{n,k,m}$. The problem in \eqref{subsubproblem} can, therefore, be
written as:
\begin{eqnarray}
&& \min_{\scriptstyle \{x_{n,k,m} \geq 0\}} \sum_{n,k,m} I_{n,k,m}\,F_{n,k,m}(I_{n,k,m},x_{n,k,m}) ~~\text{s.t.}\sum_{n,k,m} x_{n,k,m} \leq P\con, \label{eq:subproblem}
\end{eqnarray}
where $F_{n,k,m}(I_{n,k,m},x_{n,k,m})$ is defined in \eqref{defF}.
This problem is a convex optimization problem that satisfies Slater's 
condition \cite{boyd} when $x_{n,k,m} = P\con/2NKM$ 
for all $n,k,m$. Therefore,
its solution is equal to the solution of its dual problem (\ie
zero duality gap)~\cite{boyd}. To formulate the dual problem, we
write the Lagrangian of the primal problem \eqref{subproblem} as
\begin{equation}
L_{\vec{I}}(\mu, \vec{x}) = \sum_{n,k,m} I_{n,k,m}F_{n,k,m}(I_{n,k,m},x_{n,k,m})
+ \Big(\sum_{n,k,m} x_{n,k,m} - P\con \Big)\mu,
\label{eq:lagrangian}
\end{equation}
where $\mu$ is the dual variable and $\vec{x}$ is the 
$N \times K \times M$ matrix containing actual powers allocated to all 
$(n,k,m)$ combinations. Note that the Lagrangian in \eqref{lagrangian} is exactly the 
same as the Lagrangian for the CSRA problem in \eqref{clagrange}. Using \eqref{lagrangian}, the dual of the brute-force problem can be written as
\begin{equation}
\max_{\mu \geq 0} \min_{\vec{x} \succeq 0} L_{\vec{I}}(\mu, \vec{x})
=\max_{\mu \geq 0} L_{\vec{I}}(\mu, \vec{x}^*(\mu)) = L_{\vec{I}}(\mu^*_{\vec{I}}, \vec{x}^*(\mu^*_{\vec{I}})),
\end{equation}
for optimal solutions $\mu^*_{\vec{I}}$ and $\vec{x}^*(\mu^*_{\vec{I}})$.
Minimizing $L_{\vec{I}}(\mu, \vec{x})$ over $\{\vec{x} \succeq 0\}$
by equating the differential of $L_{\vec{I}}(\mu, \vec{x})$ w.r.t.\ $x_{n,k,m}$ to zero
(which is identical to the approach taken in \secref{continuous p} for the CSRA problem),
we get that, for any subchannel $n$, 
\begin{equation}
x_{n,k,m}^*(\mu) = I_{n,k,m} \, p_{n,k,m}^*(\mu).  \label{eq:pn*brute}
\vspace{-4mm}
\end{equation} 
Here,
\begin{equation}
p_{n,k,m}^*(\mu) = {\begin{cases} \tilde{p}_{n,k,m}(\mu)
	& \textrm{if}~ 0 \leq \mu \leq a_{k,m} b_{k,m} r_{k,m} 
	U'_{n,k,m}\big((1-a_{k,m})r_{k,m}\big)\E\{\gamma_{n,k}\}\\
	0 & \textrm{otherwise},
	\end{cases}} \label{eq:pn*brute2}
\end{equation}
and $\tilde{p}_{n,k,m}(\mu)$ is the unique\footnote
  {By assumption, $U'_{n,k,m}(\cdot)$ is a decreasing positive
   function and $e^{-b_{k,m} \tilde{p}_{n,k,m}(\mu) \gamma_{n,k}}$ is a strictly-decreasing 
   positive function of $\tilde{p}_{n,k,m}(\mu)$, which makes the right side of \eqref{mubrute} a 
   strictly-decreasing positive function of $\tilde{p}_{n,k,m}(\mu)$.
   }
value satisfying \eqref{gntilde}, repeated as \eqref{mubrute} for convenience.
\begin{equation}
\mu = a_{k,m} b_{k,m} r_{k,m} \E\big\{ U'_{n,k,m}\big((1-a_{k,m} 
e^{-b_{k,m} \tilde{p}_{n,k,m}(\mu)\gamma_{n,k}})r_{k,m}\big) \gamma_{n,k}e^{-b_{k,m} 
\tilde{p}_{n,k,m}(\mu) \gamma_{n,k}} \big\}. \label{eq:mubrute}
\end{equation}
Note that the Lagrangian as well as the power allocation in \eqref{lagrangian} 
and \eqref{pn*brute} are identical to that obtained for the CSRA problem in 
\eqref{clagrange} and \eqref{g_n^*}, respectively. Also recall that 
\eqref{optI_mu}-\eqref{mumax} hold even when $\vec{I}^*(\mu)$ is replaced by 
arbitrary $\vec{I}$. Thus, we have $\mu^*_{\vec{I}} \in [\mu_{\text{\sf min}}, 
\mu_{\text{\sf max}}]$, where $\mu_{\text{\sf min}}$ and 
$\mu_{\text{\sf min}}$ are defined in
\eqref{mumin} and \eqref{mumax}, respectively.

As discussed in \secref{continuous p},
$\tilde{p}_{n,k,m}(\mu)$ is a strictly-decreasing continuous function of 
$\mu$, which makes $p_{n,k,m}^*(\mu)$ a decreasing continuous function 
of $\mu$. Let us now define
\begin{equation}
X\overall^*(\vec{I},\mu) \defn \sum_{n,k,m} x_{n,k,m}^*(\mu) = \sum_{n,k,m} I_{n,k,m}p_{n,k,m}^*(\mu)
\label{eq:Xtotal}
\end{equation} 
as the total optimal power allocation for allocation $\vec{I}$ at $\mu$. 
Therefore, $X\overall^*(\vec{I},\mu)$ 
is also a decreasing continuous function of $\mu$.
This reduces our problem to finding the minimum value of 
$\mu \in [\mu_{\text{\sf min}}, 
\mu_{\text{\sf max}}]$ for which 
$X\overall^*(\vec{I},\mu) = P\con$. Such a problem structure
(\ie finding the minimum Lagrange multiplier satisfying a sum-power
constraint) yields a \emph{water-filling} solution 
(e.g., \cite{TCOM:wong:09,Feb:JSAC:Jang:03}). To obtain such a solution 
(in our case, $\mu_{\vec{I}}^*$) one can use the bisection-search algorithm given in
Table \ref{table1}. 

While there are many ways to find $\mu$, we focus on bisection-search for easy 
comparison to
the CSRA algorithm. Then, to solve the resource allocation problem for a given $\vec{I} \in 
\mc{I}\dsra$, the complexity, in terms of the number of 
times \eqref{mubrute} (or \eqref{gntilde}) is solved to yield $\hat{\mu}_{\vec{I}}$ such that 
$|\hat{\mu}_{\vec{I}} - \mu^*_{\vec{I}}| < \kappa$, is
$\big(\sum_{n,k,m}I_{n,k,m}\big) \left\lceil \log_2 \big(\frac{\mu_{\text{\sf max}} - 
\mu_{\text{\sf min}}}{\kappa}\big) \right \rceil$.
Since the brute-force algorithm examines $|\mathcal{I}\dsra| = (KM+1)^N$ 
hypotheses of $\vec{I}$, the corresponding complexity needed to find the exact 
DSRA solution is $\left \lceil \log_2 \big(\frac{\mu_{\text{\sf max}} 
- \mu_{\text{\sf min}}}{\kappa}\big)\right\rceil \times \sum_{n=1}^N n {{N}\choose{n}} (KM)^n$ or, equivalently,
\begin{eqnarray}\textstyle
\left \lceil\log_2 \big(\frac{\mu_{\text{\sf max}} 
- \mu_{\text{\sf min}}}{\kappa}\big)\right\rceil \times (KM+1)^{N-1} NKM.
\label{eq:complexitybrute}
\end{eqnarray}

Because this ``brute-force'' algorithm may be impractical to implement for 
practical values of $K$, $M$, and $N$, 
we focus, in the sequel, on lower-complexity DSRA approximations.
In doing so, we exploit insights previously gained from our study of the
CSRA problem. 

\subsection{Proposed DSRA algorithm} \label{sec:discrete imple}

Equation \eqref{Imaximin} in \secref{continuous I} demonstrated that there 
exists an optimal user-MCS allocation for the CSRA problem that
either lies in the domain of DSRA problem, i.e., $\vec{I}^*(\mu^*) \in 
\mc{I}\dsra$, or is a convex combination of two points from the domain of 
DSRA problem, i.e., $\vec{I}^*(\mu^*) = \lambda 
\vec{I}^{\scriptscriptstyle\text{\sf min}}(\mu^*) + (1-\lambda)
\vec{I}^{\scriptscriptstyle\text{\sf max}}(\mu^*)$, where 
$\vec{I}^{\scriptscriptstyle\text{\sf min}}(\mu^*) \neq 
\vec{I}^{\scriptscriptstyle\text{\sf max}}(\mu^*)$ and 
$\vec{I}^{\scriptscriptstyle\text{\sf min}}(\mu^*), 
\vec{I}^{\scriptscriptstyle\text{\sf max}}(\mu^*) \in 
\mc{I}\dsra$.
(Note that if $\vec{I}\in \mc{I}\csra$ and
$\vec{I}\in \{0, 1\}^{N \times K \times M}$, then $\vec{I}\in \mc{I}\dsra$.)
This observation motivates us to attack the DSRA problem using the CSRA 
algorithm. 
In this section, we provide the details of such an approach.

The following lemma will be instrumental in understanding the relationship 
between the CSRA and DSRA problems and will serve as the basis for allocating
resources in the DSRA problem setup.
\begin{lemma} \label{lem:lem7}
If the solution of the Lagrangian dual of the CSRA problem \eqref{cdual} for 
a given $\mu$ 
is such that $\vec{I}^*(\mu) \in \{0, 1\}^{N \times K \times M}$, and the
corresponding total power is $X^*\overall(\mu)$ as in \eqref{xoverall}, then
the solution to the optimization problem
\begin{eqnarray*}
(\mathbb{P}^*, \mathbb{I}^*) &=& \argmax_{\substack{\scriptstyle \{\mathbb{P} \succeq 0\}\\[0.8 mm]
	\scriptstyle \vec{I} \in \mc{I}\dsra}} \sum_{n,k,m} \mathbb{I}_{n,k,m}
	\E\Big\{ U_{n,k,m}\big((1-a_{k,m} e^{-b_{k,m} \mathbb{P}_{n,k,m} \gamma_{n,k}})
	r_{k,m}\big) \Big\} 
 ~\mathrm{s.t.}\sum_{n,k,m} \mathbb{I}_{n,k,m} \mathbb{P}_{n,k,m}\leq
X^*\overall(\mu) 
\\[-5mm]
\end{eqnarray*}
satisfies $\mathbb{I}^* = \vec{I}^*(\mu)$ and, for every $(n,k,m)$,
$\displaystyle 
\mathbb{P}_{n,k,m}^* = {\begin{cases} \frac{x_{n,k,m}^*(\mu, \vec{I}^*(\mu))}{I_{n,k,m}^*(\mu)} &
\textrm{if}~ I_{n,k,m}^*(\mu) \neq 0 \\
0 & \textrm{otherwise}. \end{cases}}$ 
\end{lemma}
\begin{IEEEproof}
A proof sketch is given in Appendix \ref{appendix8}. 
For the full proof, see \cite{mythesis}.
\end{IEEEproof}

From the above lemma, we conclude that if a $\mu$ exists such that
$\vec{I}^*(\mu) \in \mc{I}\dsra$ and 
$X^*\overall(\mu) = P\con$, then
the DSRA problem is solved exactly by the CSRA solution 
$(\vec{I}^*(\mu), \vec{x}^*(\mu, \vec{I}^*(\mu)))$, i.e., the optimal 
user-MCS allocation $\vec{I}^*\dsra$ equals $\vec{I}^*(\mu)$ and the 
optimal power allocation, 
$\vec{p}^*\dsra$, for any $(n,k,m)$, is 
\begin{equation}
p_{n,k,m,\, \text{\sf DSRA}}^* = {\begin{cases} \frac{x_{n,k,m}^*(\mu, \vec{I}^*(\mu)))}{I_{n,k,m}^*(\mu)} & \textrm{if}~
I_{n,k,m}^*(\mu) \neq 0 \\
0 & \textrm{otherwise}. \end{cases}}
\end{equation}
Recall that the 
optimal total power achieved for a given value of Lagrange multiplier 
$\mu$, i.e., $X\overall^*(\mu) = \sum_{n,k,m}x_{n,k,m}^*(\mu, 
\vec{I}^*(\mu))$, is piece-wise continuous and that a discontinuity (or 
``gap'') occurs at $\mu$ when multiple allocations achieving the same optimal 
value of Lagrangian exist. When the sum-power constraint,
$P\con$, lies in one of those ``gaps,'' the optimal allocation for the
CSRA problem equals a convex combination of two elements from the
set $\mc{I}\dsra$, and the CSRA solution is not admissible for DSRA. 
In such cases, we are motivated to choose the approximate DSRA solution
$\vec{\hat{I}}\dsra \in \{ \vec{I}^{\scriptscriptstyle\text{\sf min}}(\mu), 
\vec{I}^{\scriptscriptstyle\text{\sf max}}(\mu)\}$ 
yielding highest utility.
In Table \ref{table1}, we detail an implementation of our proposed 
DSRA algorithm that has significantly
lower complexity than brute-force. 
The numerical simulations in \secref{simulations} show that 
its performance is very close to optimal.
Moreover, the following lemma bounds the asymptotic difference in utility 
achieved by the exact DSRA solution and that produced by our proposed 
DSRA algorithm.

\begin{lemma} \label{lem:lem8}
Let $\mu^*$ be the optimal $\mu$ for the CSRA problem and $\underline{\mu}, \bar{\mu}$ 
be such that $\mu^* \in [\underline{\mu}, \bar{\mu}]$. Let $U^*\dsra$ and 
$\hat{U}\dsra(\underline{\mu},\bar{\mu})$ be the utilities achieved by 
the exact DSRA solution and the proposed DSRA algorithm, respectively. Then,
\begin{eqnarray}
0 ~\leq~ U^*\dsra - \lim_{\underline{\mu} \to \bar{\mu}} 
\hat{U}\dsra(\underline{\mu},\bar{\mu}) 
&\leq& (\mu^* - \mu_{\text{\sf min}}) 
\big(P\con - X\overall^*(\vec{I}^{\text{\sf min}}(\mu^*), \mu^*)\big) \label{eq:DSRAbound} \\
&\leq& {\begin{cases} 0 & \textrm{if}~ |S_n(\mu^*)| \leq 1~\forall n \\
\big(\mu_{\text{\sf max}} - \mu_{\text{\sf min}}\big)P\con
 & \textrm{otherwise} \end{cases}}. \label{eq:bound}
\end{eqnarray}
\end{lemma}
\begin{IEEEproof}
The proof is given in Appendix \ref{appendix9}.
\end{IEEEproof}
It is interesting to note that the bound \eqref{bound} does
not scale with number of users $K$ or subchannels $N$.

The complexity of the proposed DSRA algorithm is marginally greater than 
that of the CSRA algorithm, since an additional comparison of two possible 
user-MCS allocation choices is involved. In units of solving \eqref{gntilde} 
for a given $(n,k,m,\mu)$, the DSRA complexity is at most 
\begin{equation}\textstyle
N(KM+2)\left\lceil \log_2 \big(\frac{\mu_{\text{\sf max}} - 
\mu_{\text{\sf min}}}{\kappa}\big) \right \rceil.
\label{eq:complexitydsra}
\end{equation}
Comparing \eqref{complexitybrute} and \eqref{complexitydsra}, we find that 
the complexity of the proposed DSRA algorithm is polynomial in 
$N, K, M$, which is considerably less than that of the brute-force 
algorithm (i.e., exponential in $N$).

\subsection{Discussion}   \label{sec:discussion}

Before concluding this section, we make some remarks about our approach 
to DSRA and its connections to CSRA.
First, we note that the DSRA problem is an integer-programming problem due
to the discrete domain $\{0,1\}$ assumed for $I_{n,k,m}$.
Because integer programming problems are generally NP-hard (recall
our ``brute force'' DSRA solution), one is strongly motivated to find a
polynomial-complexity method whose performance is as high as possible.
One possible approach is based on ``relaxation,'' whereby 
the discrete domain is relaxed to an interval domain, 
the relaxed problem is solved (with polynomial complexity), 
and the obtained solution is mapped back to the discrete domain.
In fact, relaxation was previously employed for OFDMA frequency-scheduling 
in \cite{Oct:JSAC:Wong:99,TCOM:wong:09}, and the DSRA approximation that we 
propose in \secref{discrete imple} can also be interpreted as a form 
of relaxation.

The optimization literature suggests that relaxation is successful 
in some---but not all---cases, implying that relaxation-based OFDMA 
algorithms must be designed with care.
For example, relaxation has widely used to solve \emph{linear} integer 
programs (LIPs) \cite{Benichou.Gauthier.ea:71,Lodi:2010,Cornuejols.et.al:2010}.
The DSRA, however, is a mixed-integer nonlinear program (MINLP),
and for such problems relaxation does not always perform well
\cite{Jeroslow:73,Cornuejols.et.al:2010}.
Now, one could cite the analysis in \cite[p.\ 371]{bertsekas}, which shows 
that---for a broad class of integer programming problems---the duality gap 
goes to zero as the number of integer variables goes to infinity, to 
suggest that the DSRA problem can be well approximated by its relaxed
counterpart, CSRA, as the number of OFDMA subchannels $N\rightarrow\infty$.
However, in practice, the number of subchannels $N$ is often quite small,
preventing the application of this argument.
For example, in LTE systems \cite{LTE,LTE2}, each subchannel consists of 
$12$ subcarriers, so that only $25$ subchannels are used for
$5$ MHz bandwidths, and only $6$ are used for $1.4$ MHz bandwidths. 

The above considerations have motivated us to investigate, in detail, 
the relationship between the continuous and discrete resource allocation 
scenarios. 
The results of our investigation include insights into the dissimilarity 
between CSRA and DSRA solutions (e.g., \lemref{lem1} and \lemref{lem7}),
and an efficient polynomial-complexity DSRA approximation that
(as we shall see in \secref{simulations}) performs near-optimally for all 
$N$ and admits the tight performance bound \eqref{bound}.

\section{Numerical Evaluation} \label{sec:simulations}
\newcommand{\pilot}{_{\textsf{pilot}}}
\newcommand{\eye}{\textbf{I}}

In this section, we analyze the performance of an OFDMA downlink 
system that uses the proposed CSRA and DSRA algorithms for scheduling 
and resource allocation under different system parameters.
Unless otherwise specified, we use the sum-goodput utility 
$U_{n,k,m}(g) = g$.

For downlink transmission, the BS employs an uncoded $2^{m+1}$-QAM signaling 
scheme with MCS index $m \in \{1, \ldots, 15\}$.
In this case, we have $r_{k,m}=m+1$ bits per symbol and one symbol per codeword.
In the error rate model
$\epsilon_{k,m}(p \gamma) = a_{k,m} e^{-b_{k,m} p \gamma}$, we choose $a_{k,m} = 1$
and $b_{k,m} = 1.5/(2^{m+1}-1)$ because the actual
symbol error rate of a $2^{m+1}$-QAM system is proportional to 
$\exp(-1.5 p \gamma/(2^{m+1}-1))$ in the high-$(p \gamma)$ regime \cite{Proakis:Book:08}
and is $\approx 1$ when $p \gamma= 0$. 
We use the standard OFDM model \cite{VTC:beek:95}
to describe the (instantaneous) frequency-domain observation made by the 
$k^{th}$ user on the $n^{th}$ subchannel:
\begin{equation}
y_{n,k} = h_{n,k} x_{n} + \nu_{n,k}, ~~ 
	\text{for $n \in \{1, \ldots, N\}$ and $k\in\{1,\dots,K\}$}
					\label{eq:ynk}
\end{equation}
In \eqref{ynk},
$x_{n}$ denotes the QAM symbol broadcast by the BS on the $n^{th}$ subchannel,
$h_{n,k}$ the gain of the $n^{th}$ subchannel between the $k^{th}$ user and the BS,
and $\nu_{n,k}$ a corresponding complex Gaussian noise sample.
We assume that $\{\nu_{n,k}\}$ is unit variance and white across $(n,k)$, and we
recall that the exogenous subchannel-SNR satisfies $\gamma_{n,k} = |h_{n,k}|^2$.
We furthermore assume that the $k^{th}$ user's frequency-domain channel gains
$\vec{h}_k=(h_{1,k}, \ldots, h_{N,k})^T\in\Complex^N$ are related to its
channel impulse response $\vec{g}_k=(g_{1,k}, \ldots, g_{L,k})^T\in\Complex^L$
via $\vec{h}_k=\vec{F}\vec{g}_k$, where $\vec{F}\in\Complex^{N\times L}$
contains the first $L(<N)$ columns of the $N$-DFT matrix, and
where $\{g_{l,k}\}$ are i.i.d. over $(l,k)$ and drawn
from a zero-mean complex Gaussian distribution with variance $\sigma_g^2$
chosen so that $\E\{\gamma_{n,k}\}=1$. Since the total 
available power for all subchannels at the base-station is $P\con$, the
average available SNR per subchannel will be denoted by 
$\textsf{SNR} = \frac{P\con}{N}\E\{\gamma_{n,k}\}$.

To model imperfect CSI, we assume that there is a channel-estimation period
during which the mobiles take turns to each broadcast one pilot OFDM symbol,
from which the BS estimates the corresponding subchannel gains.
Furthermore, we assume that the channels do not vary between pilot and data periods.
To estimate $\vec{h}_k$, we assume that the BS observes 
$\tvec{y}_k=\sqrt{p\pilot}\,\vec{h}_k+\tvec{\nu}_k\in\Complex^N$. 
Note that the average SNR per subchannel under pilot transmission is
$\textsf{SNR}\pilot = p\pilot\E\{\gamma_{n,k}\}$.
The channel $\vec{h}_k$ and the
pilot observations $\tvec{y}_k$ are zero-mean jointly Gaussian, and furthermore
$\vec{h}_k|\tvec{y}_k$ is Gaussian with mean $\E\{\vec{h}_k|\tvec{y}_k\}
=\vec{R}_{\vec{h}_k,\tvec{y}_k}\vec{R}_{\tvec{y}_k,\tvec{y}_k}^{-1}\tvec{y}_k$
and covariance $\cov(\vec{h}_k|\tvec{y}_k)=\vec{R}_{\vec{h}_k,\vec{h}_k}
-\vec{R}_{\vec{h}_k,\tvec{y}_k}\vec{R}_{\tvec{y}_k,\tvec{y}_k}^{-1}
\vec{R}_{\tvec{y}_k\vec{h}_k}$, where
$\vec{R}_{\vec{z}_1, \vec{z}_2}$ denotes the cross-correlation
of random vectors $\vec{z}_1$ and $\vec{z}_2$ \cite[pp. $155$]{poor}.
Since
$\vec{R}_{\vec{h}_k,\vec{h}_k}=\sigma_g^2\vec{FF}'$,
$\vec{R}_{\vec{h}_k,\tvec{y}_k}=\sqrt{p\pilot}\sigma_g^2\vec{FF}'$,
and
$\vec{R}_{\tvec{y}_k,\tvec{y}_k}=p\pilot\sigma_g^2\vec{FF}'+\eye$
(where $\eye$ denotes the identity matrix),
it is straightforward to show that the elements on the diagonal
of $\cov(\vec{h}_k|\tvec{y}_k)$ are equal.
Furthermore, $\E\{\vec{h}_k|\tvec{y}_k\}$ can be recognized as the
pilot-aided MMSE estimate of $\vec{h}_k$.
In summary, conditioned on the pilot observations, $h_{n,k}$ is Gaussian
with mean $\hat{h}_{n,k}$ given by the $n^{th}$ element of
$\E\{\vec{h}_k|\tvec{y}_k\}$, and with variance $\sigma_e^2$
given by the first diagonal element of $\cov(\vec{h}_k|\tvec{y}_k)$.
Thus, conditioned on the pilot observations, $\gamma_{n,k}$
has a non-central chi-squared distribution with two degrees of freedom.

We will refer to the proposed CSRA and DSRA algorithms implemented under
imperfect CSI as ``CSRA-ICSI'' and ``DSRA-ICSI,'' respectively.
Their performances will be compared to that of ``CSRA-PCSI,'' i.e., CSRA
implemented under perfect CSI,
which serves as a performance upper bound, and
\emph{fixed-power random-user scheduling} (FP-RUS), which serves as a
performance lower bound.
FP-RUS schedules, on each subchannel, one user selected uniformly from
$\{1,\ldots,K\}$, to which it allocates power $P\con/N$ and the fixed
MCS $m$ that maximizes expected goodput. Unless specified, the 
number of OFDM subchannels is $N = 64$, the number 
of users is $K = 16$, the impulse response length is $L=2$,
the average SNR per subchannel is $\textsf{SNR}=10$ dB, the pilot SNR is
$\textsf{SNR}\pilot = -10$ dB, and the DSRA/CSRA tuning parameter is
$\kappa=0.3/P\con$ (recall Table \ref{table1}).
In all plots, goodput values were empirically averaged
over $1000$ realizations.

\begin{figure*}[t]
\renewcommand{\baselinestretch}{1.0}
\newcommand{\pwid}{3in}
\newcommand{\fwid}{3.2in}
\newcommand{\hsp}{2mm}
  \begin{minipage}{\pwid}
  \hspace{\hsp}
    \psfrag{mu}[t][][0.8]{\sf $\mu$}
	\psfrag{optX}[][][0.7]{\sf $X\overall^*(\mu)$}
	\psfrag{optL}[][][0.7]{\sf $L(\mu, \vec{I}^*(\mu), \vec{x}^*(\mu, \vec{I}^*(\mu)))$}
    \includegraphics[width=\fwid]{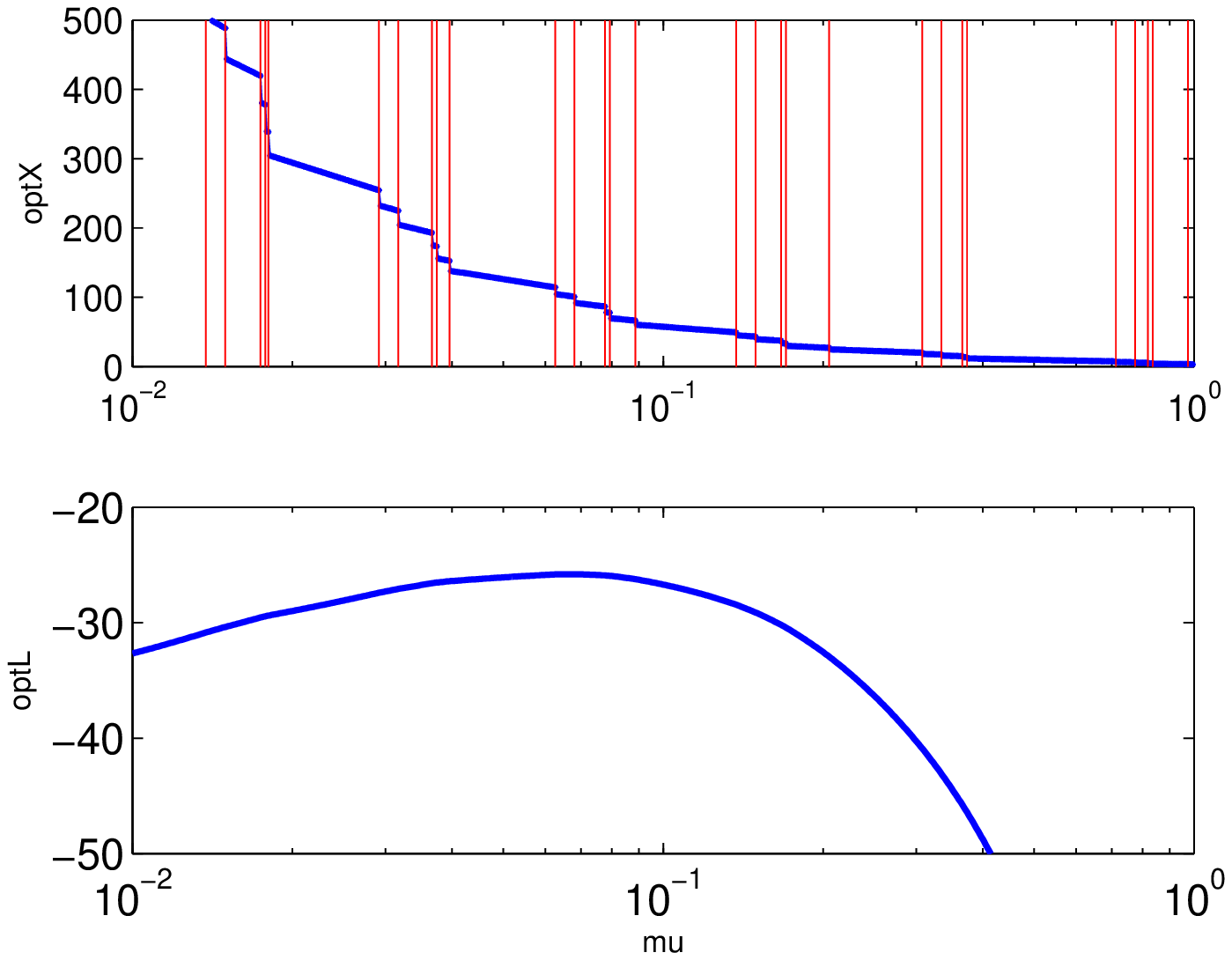}
    \caption[hang]{\footnotesize Prototypical plot of 
	$X^*\overall(\mu)$ and $L(\mu, \vec{I}^*(\mu), \vec{x}^*(\mu, 
	\vec{I}^*(\mu)))$ as a function of $\mu$ for $N=K=5$, and $P\con = 100$. 
	(See \secref{simulations} for details.) The red vertical 
	lines in the top plot show that a 
	change in $\vec{I}^*(\mu)$ occurs at that $\mu$.}
    \label{fig:sump}
  \end{minipage}
  \hspace{0.6cm}  \hspace{\hsp}  \begin{minipage}{\pwid}
    \psfrag{pilots}[t][][0.7]{\sf $\textsf{SNR}\pilot$ (in dB)}
	\psfrag{utility per subchannel}[][][0.7]{\sf Goodput (bpcu)}
    \includegraphics[width=\fwid]{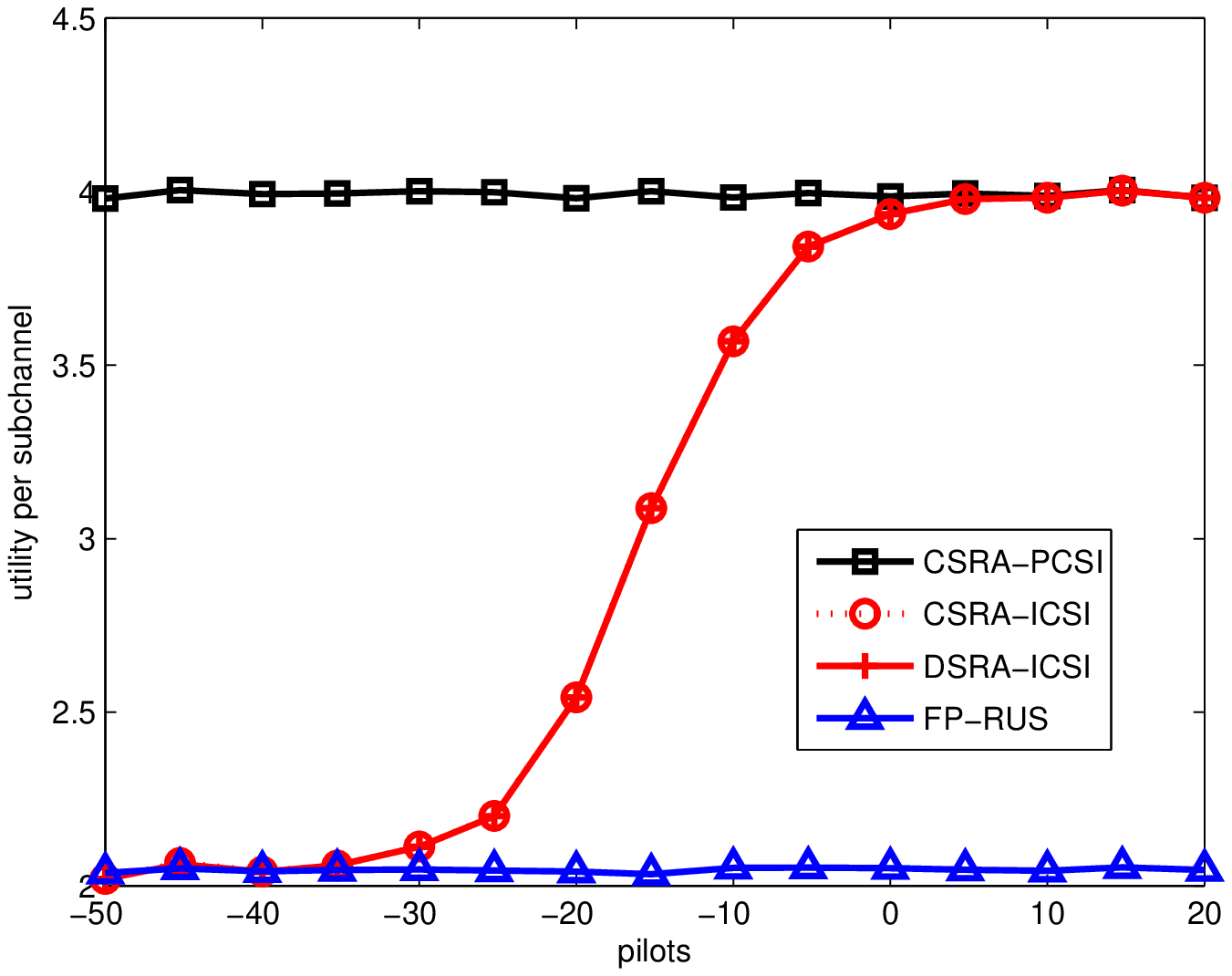}
    \caption[hang]{\footnotesize Average goodput per subchannel versus $\textsf{SNR}\pilot$. Here, 
	$N=64$, $K=16$, and $\textsf{SNR} = 10$ dB.}
    \label{fig:withpilots2}
  \end{minipage}
\vspace{-5mm}
\end{figure*}

\Figref{withpilots2} plots the subchannel-averaged
goodput achieved by the above-described scheduling and resource-allocation 
schemes for
different grades of CSI. In this curve, $\textsf{SNR}\pilot$ is varied so as to 
obtain estimates of subchannel SNR with different grades of accuracy. All other 
parameters remain unchanged. The plot shows that, as
$\textsf{SNR}\pilot$ is increased, the performance of the proposed schemes
(under the availability of imperfect CSI) increases from that of FP-RUS to
that achieved by CSRA-PCSI. This is expected because, with increasing 
$\textsf{SNR}\pilot$, the BS uses more
accurate channel-state information for scheduling and resource allocation,
and thus achieves higher goodput. The plot also shows that, even though
the proposed CSRA algorithm exactly solves the CSRA problem and the
proposed DSRA algorithm approximately solves the DSRA problem, their
performances almost coincide. In particular, 
although the goodput achieved by CSRA-ICSI scheme exceeded that of
DSRA-ICSI scheme in up-to $49\%$ of the realizations,
the maximum difference in the subchannel-averaged goodput was merely 
$4 \times 10^{-3}$ bits per channel-use (bpcu). Since the 
DSRA-ICSI schemes cannot achieve a
sum-goodput higher than that achieved by the CSRA-ICSI scheme,
it can be deduced that the proposed DSRA algorithm is exhibiting
near-optimal performance.


\begin{figure*}[t]
\renewcommand{\baselinestretch}{1.0}
\newcommand{\pwid}{3in}
\newcommand{\fwid}{3.2in}
\newcommand{\hsp}{2mm}
  \begin{minipage}{\pwid}
    \psfrag{users}[t][][0.7]{\sf Number of users, $K$}
	\psfrag{utility per subchannel}[][][0.7]{\sf Goodput (bpcu)}
    \includegraphics[width=\fwid]{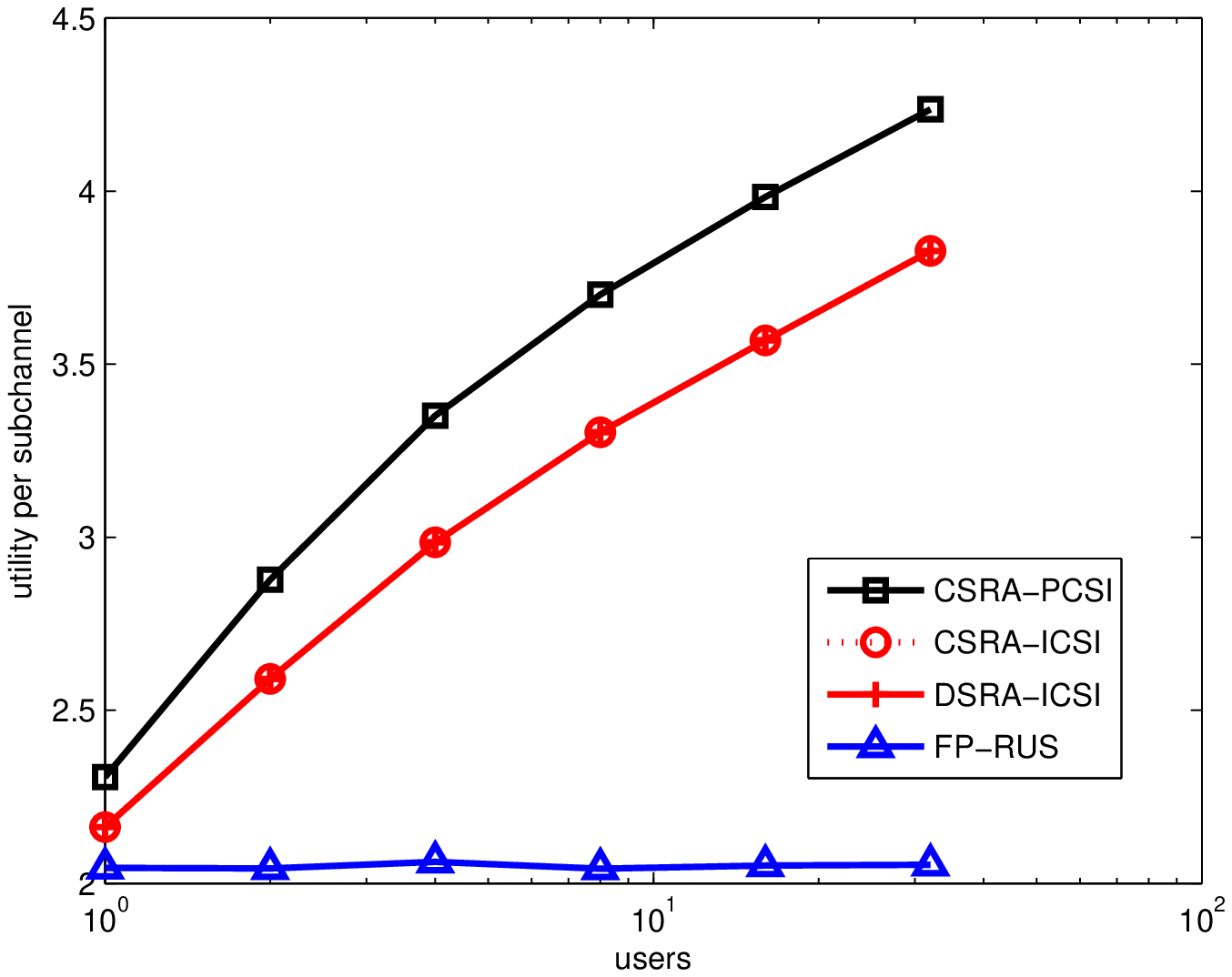}
    \caption[hang]{\footnotesize Average goodput per subchannel versus number of users, $K$. 
	In this plot, $N = 64$, 
	$\textsf{SNR} = 10$ dB, and $\textsf{SNR}\pilot = -10$ dB.}
    \label{fig:versus_K}
  \end{minipage}
  \hspace{0.6cm}
  \hspace{\hsp}
  \begin{minipage}{\pwid}
  \hspace{\hsp}
    \psfrag{Pcon}[t][][0.7]{\sf SNR (in dB)}
	\psfrag{utility per subchannel}[][][0.7]{\textsf{Goodput (bpcu)}}
	\psfrag{Average Bound}[][][0.7]{\textsf{Goodput (bpcu)}}
    \includegraphics[width=\fwid]{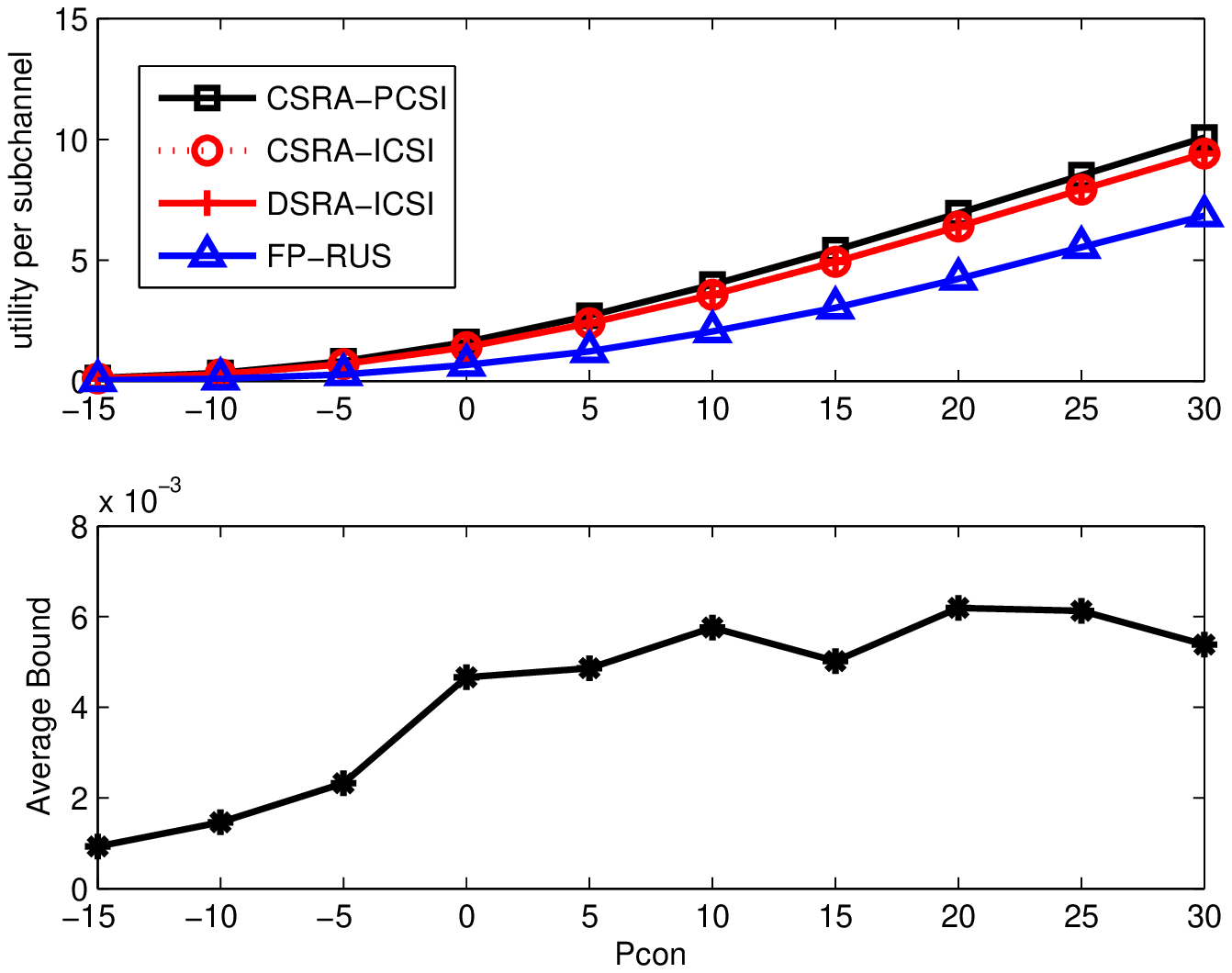}
    \caption[hang]{\footnotesize The top plot shows the average goodput per subchannel 
    	as a function of $\textsf{SNR}$. The 
	bottom plot shows the average bound on the optimality gap between 
	the proposed and exact 
	DSRA solutions (given in \eqref{DSRAbound}), i.e., the average 
	value of $(\mu^* - \mu_{\text{min}}) (P\con - 
	X\overall^*(\vec{I}^{\scriptscriptstyle\text{min}}, \mu^*))/N$. In 
	this plot, $N=64$, $K=16$, and $\textsf{SNR}\pilot = -10$ dB.}
    \label{fig:versus_Pcon}
  \end{minipage}

\vspace{-5mm}
\end{figure*}

\Figref{versus_K} plots the subchannel-averaged goodput versus the number 
of available users, $K$, ranging between $1$ and $32$. It shows 
that, as $K$ increases, the goodput per subchannel achieved by the 
proposed schemes increase under both perfect and imperfect CSI,
whereas that achieved by the FP-RUS scheme 
remains constant. This is because, in the former 
case, the availability of more users can be exploited 
to schedule users with stronger subchannels, whereas, in the 
FP-RUS scheme, users are scheduled without regard to 
the instantaneous channel conditions. Similar to the
observations in the previous plots, the performance difference
between the proposed CSRA and DSRA algorithms remains negligible.
In particular, although the goodput achieved by CSRA-ICSI 
exceeded that of DSRA-ICSI in up-to $29\%$ of the 
realizations, the maximum difference in the subchannel-averaged goodput 
was merely $7 \times 10^{-4}$ bpcu.

In \Figref{versus_Pcon}, the top plot shows the subchannel-averaged goodput 
and the bottom plot shows the subchannel and realization-averaged value of the bound 
\eqref{DSRAbound} on the DSRA-ICSI optimality gap as a function 
of $\textsf{SNR}$. In the top plot, it can be seen that, as 
$\textsf{SNR}$ increases, the difference between CSRA-PCSI and CSRA-ICSI (or, DSRA-ICSI) 
increases. However, the difference grows slower than the difference 
between CSRA-PCSI and FP-RUS. Interestingly, even for high values of 
$\textsf{SNR}$, the performance of CSRA-ICSI and DSRA-ICSI remain almost identical. 
In particular, although the goodput achieved by CSRA-ICSI scheme exceeded 
that of DSRA-ICSI scheme in up-to $28\%$ of the realizations,
the maximum difference in the subchannel-averaged goodput was merely 
$4 \times 10^{-5}$ bpcu. The bottom plot, which illustrates
the average value of $(\mu^* - \mu_{\text{\sf min}})\big(P\con - 
X\overall^*(\vec{I}^{\scriptscriptstyle\text{\sf min}}, \mu^*)\big)$ 
over all realizations and subchannels w.r.t.\ $\textsf{SNR}$, shows 
that the loss in sum-goodput over all subchannels due to the
sub-optimality of proposed DSRA solution under imperfect CSI is bounded by
$7 \times 10^{-3}$ bpcu, even when the subchannel-averaged goodput
of DSRA-ICSI is of the order of tens of bpcu.
These results confirm that the bound \eqref{DSRAbound} is quite tight at high 
$\textsf{SNR}$.

\begin{figure*}[t]
\renewcommand{\baselinestretch}{1.0}
\newcommand{\pwid}{3in}
\newcommand{\fwid}{3.2in}
\newcommand{\hsp}{2mm}

  \begin{minipage}{\pwid}
   \psfrag{Bandwidth}[][][0.7]{\sf Bandwidth $p\gamma$}
   \psfrag{utility function}[][][0.7]{\sf Utility}
   \psfrag{Goodput}[][][0.7]{\sf goodput}
   \psfrag{SNR}[t][][0.7]{\sf SNR (in dB)}
	\psfrag{Utility}[][][0.7]{\textsf{Sum utility}}
	\psfrag{sum utility}[][][0.7]{\textsf{Sum utility}}
	\psfrag{weight}[][][0.7]{\textsf{weight} $w_1$}
	\psfrag{class I sum-utility}[l][l][0.45]{\textsf{Sum Utility--Class 1}}
	\psfrag{class II sum-utility}[l][l][0.45]{\textsf{Sum Utility--Class 2}}
	\psfrag{Naive sum utility}[l][l][0.45]{\sf Sum Utility--Naive}
	\psfrag{sum-utility}[l][l][0.45]{\sf Sum Utility}
    \includegraphics[width=\fwid]{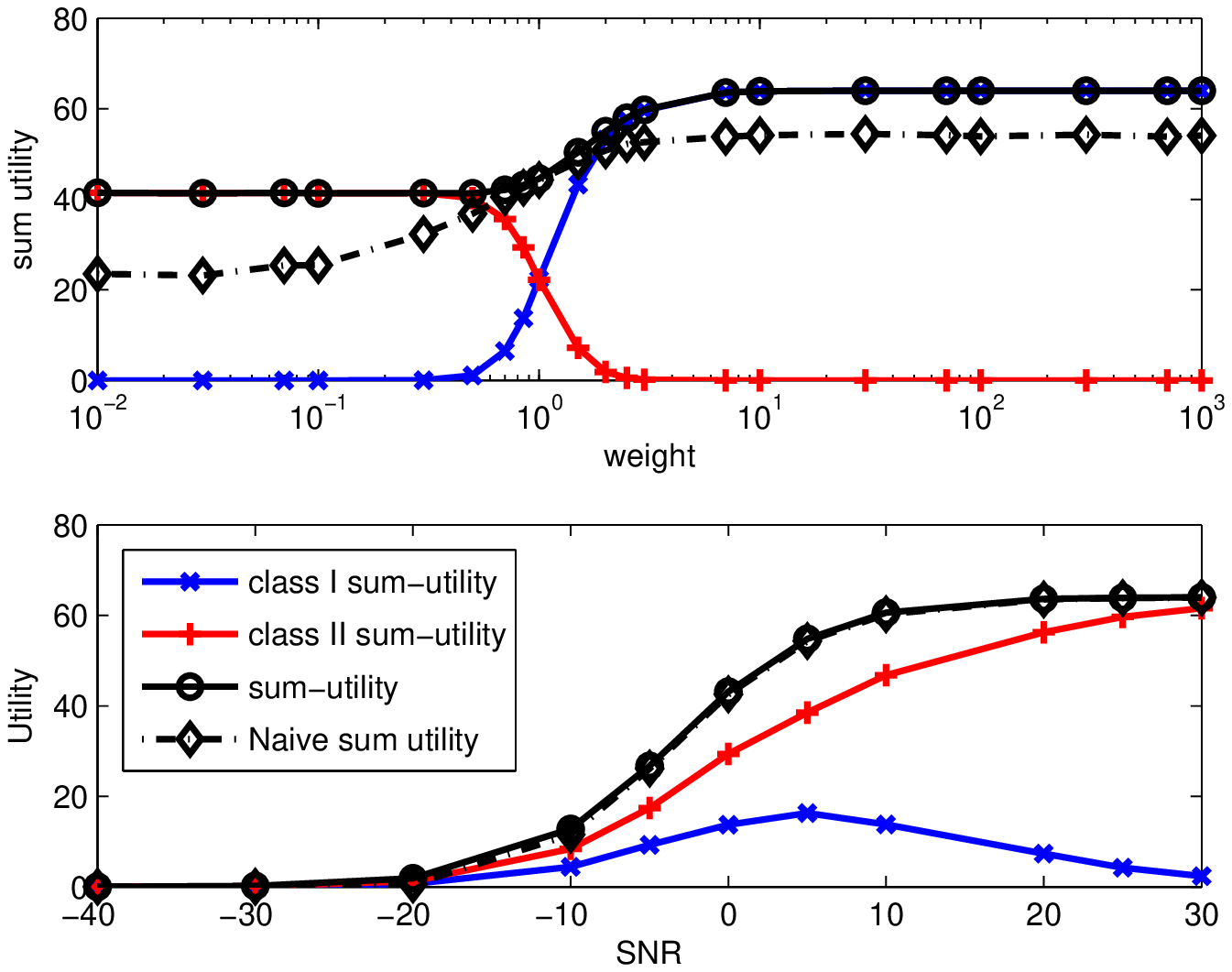}
    \caption[hang]{\footnotesize The top plot shows sum utility versus $w_1$
	when $w_2=1$, $\textsf{SNR}=0$ dB. 
	The bottom plot shows the sum-utility versus $\textsf{SNR}$ when $w_1=0.85, w_2 =1$. 
	Here, $N=64$, $K=16$, and $\textsf{SNR}\pilot = -10$ dB.}
    \label{fig:versus_weight_k}
  \end{minipage}
  \hspace{0.6cm}
  \hspace{\hsp}
  \begin{minipage}{\pwid}
  \hspace{\hsp}
        \psfrag{Number of iterations}[t][][0.7]{\sf Number of $\mu$-updates}
	\psfrag{Total utility}[][][0.7]{\textsf{Sum utility (bpcu)}}
	\psfrag{Mean Deviation}[][][0.7]{\textsf{Mean deviation}}
	\psfrag{DSRA}[l][l][0.45]{\textsf{DSRA}}
	\psfrag{CSRA}[l][l][0.45]{\textsf{CSRA}}
	\psfrag{[9]}[l][l][0.45]{\sf [9]}
	\psfrag{[7]}[l][l][0.45]{\sf [7]}
    \includegraphics[width=\fwid]{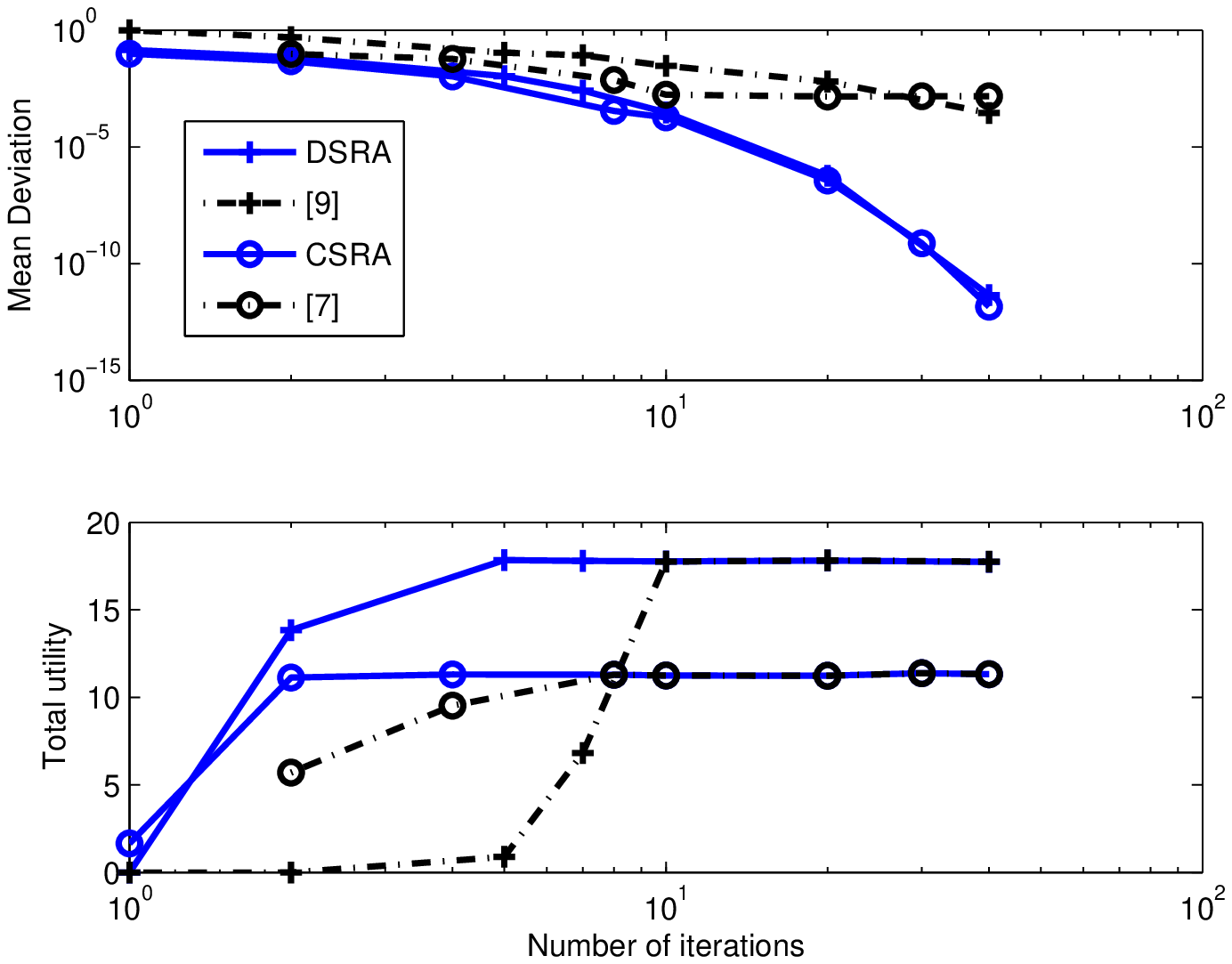}
    \caption[hang]{\footnotesize The top plot shows the mean deviation of the estimated 
    	dual variable $\mu$ from $\mu^*$, and the bottom plot shows average sum-utility,
	as a function of the number of $\mu$-updates. Here, $N=64$, $K = 16$, 
	$\textsf{SNR} = 10$ dB, and $\textsf{SNR}\pilot = -10$ dB.}
    \label{fig:versus_5_10}
  \end{minipage}
\vspace{-5mm}
\end{figure*}


{%
\Figref{versus_weight_k} shows the performance of the proposed DSRA algorithm under 
a sum-utility criterion that is motivated by a common pricing model for 
an elastic application such as file-transfer \cite{Pricingsurvey2,shenker}.
In particular, we partitioned the $K=16$ users into two classes:
$k\in\{1,\dots,8\}\defn\mc{K}_1$ is ``Class 1''  and
$k\in\{9,\dots,16\}\defn\mc{K}_2$ is ``Class 2,''
and we ran DSRA with the utility 
$U_{k}(g) \defn (1-e^{-w_1 g}) \mathbf{1}_{k\in \mc{K}_1} + 
(1-e^{-w_2 g}) \mathbf{1}_{k\in \mc{K}_2}$, where $\mathbf{1}_{\mc{E}}$ denotes 
the indicator of event $\mc{E}$.
The utility can be regarded as the revenue earned by the operator: 
when $w_i>w_j$, Class-$i$ users pay more (for a given goodput $g$) than Class-$j$ 
users in exchange for priority service. 
In \figref{versus_weight_k}, we show the resulting DSRA-maximized
utility summed over all users, as well as that summed over each individual user class.
For comparison, we show the utility (summed over all users) when DSRA is ``naively'' used 
to maximize sum-\emph{goodput} instead of sum-utility.
The top plot in \figref{versus_weight_k} shows performance as a function of $w_1$, 
for fixed $w_2=1$ and $\textsf{SNR}=0$ dB. 
There the behavior is as expected: when $w_1\ll w_2=1$ (i.e., Class-1 users pay much
less) DSRA allocates the overwhelming majority of the resources to Class-2 users, in an
effort to earn more revenue.
Meanwhile, when $w_1\gg w_2=1$, the overwhelming majority of resources are allocated
to Class-1 users. 
Moreover, it is evident that the naive goodput-maximizing scheme does not earn the operator 
as much revenue as the utility-maximizing scheme (outside of the trivial case that $w_1=w_2$).
The bottom plot in \figref{versus_weight_k} shows the above described sum-utilities 
as a function of \textsf{SNR}, for fixed $w_1=0.85$ and $w_2=1$.
There it can be seen that, at low \textsf{SNR}, the two classes achieve
proportional utilities while, at high \textsf{SNR}, the utility of
Class-1 users tend to zero. 
This behavior can be explained as follows:
At low \textsf{SNR}, the goodputs $g$ are small, in which case 
$1-e^{-w_i g}\approx w_i g$, so that
$U_k(g) \approx w_i g \mathbf{1}_{k\in\mc{K}_i}$, i.e., 
weighted-goodput utility. 
At high \textsf{SNR}, this approximation does not hold because the goodputs 
$g$ are usually large,
and this particular pricing-based utility becomes increasingly unfair.
}

%


{%
In \Figref{versus_5_10}, we compare the performances of our proposed 
algorithms to the state-of-the-art algorithms in \cite{TCOM:wong:09,berry:tcom:2009}. 
In particular, we first compare 
the golden-section-search based algorithm from \cite{berry:tcom:2009} to
our CSRA algorithm.
For CSRA, we choose the utility function and the SNR distributions to maximize the
upper bound on capacity computed via the effective SNR 
$\frac{1}{K} \sum_{n,k}\log 
\big(1+\frac{p_{n,k,1}|\hat{h}_{n,k}|^4}{|\hat{h}_{n,k}|^2+  \sigma_e^2 
\, p_{n,k,1}|\hat{h}_{n,k}|^2}\big)$
from \cite[Eq.\ (4)]{berry:tcom:2009}.
Second, we compare the subgradient-based algorithm proposed for discrete 
allocation in \cite{TCOM:wong:09} to our DSRA algorithm.
For DSRA, we choose the utility $U_{n,k,m}(g) = \frac{1}{K}\log(1-\log(1-g))~\forall n,k,m$,
so that we maximize $\frac{1}{K}\sum_{n,k}\E\{\log (1+p_{n,k,1}\gamma_{n,k})\}$,
as in \cite{TCOM:wong:09}.
The top plot in \figref{versus_5_10} shows the mean deviation of the estimated value 
of the dual variable $\mu$ from the optimum (i.e., $\mu^*$), and the bottom plot 
shows the total utility achieved as a function of the number of
$\mu$-updates.
For the subgradient-based algorithm in~\cite{TCOM:wong:09}, we set the 
step-size in the $i^{\textrm{th}}$ $\mu$-update to be $1/i$. 
In the top plot, it can be seen that the proposed algorithms
outperform the algorithms in \cite{TCOM:wong:09,berry:tcom:2009} and converge toward
$\mu^*$ at a much faster rate. 
The bottom plot shows that the proposed algorithms achieve a much higher utility 
than the algorithms in \cite{TCOM:wong:09,berry:tcom:2009} for the first few 
$\mu$-updates, illustrating the speed of our approaches.
Note that the golden-section algorithm only provides estimates of $\mu^*$ at even 
numbers of $\mu$-updates.}



\section{Conclusion}				\label{sec:conclusion}

In this paper, we considered the problem of joint scheduling
and resource allocation (SRA) in downlink OFDMA systems under
imperfect channel-state information. 
We considered two scenarios: 1) when subchannel sharing is allowed, 
and 2) when it is not. 
Both cases were framed as optimization problems that maximize a utility 
function subject to a sum-power constraint. 
Although the optimization problem in the first scenario (the so-called 
``continuous'' or CSRA case) was found to be non-convex, 
we showed that it can be converted to a convex optimization 
problem and solved using a dual optimization approach with zero duality gap. 
An algorithmic implementation of the CSRA solution was also provided. 
The optimization problem faced in the second scenario (the so-called 
``discrete'' or DSRA case) was found to be a mixed-integer programming problem. 
To attack it, we linked the DSRA problem to the CSRA problem,
and showed that, in some cases, the DSRA solution coincides with 
the CSRA solution. 
For the case that the solutions do not coincide, we proposed a practical
DSRA algorithm and bounded its performance.
Numerical results were then presented under a variety of settings.
The performance of the proposed CSRA and DSRA algorithms schemes under imperfect 
CSI were compared to those under perfect CSI and no instantaneous
CSI (i.e., fixed-power random scheduling). 
In all cases, it was found that the proposed imperfect-CSI-based algorithms 
offer a significant advantage over schemes that do not use instantaneous CSI.
{
Next, our DSRA bound was numerically evaluated and found to be extremely tight.
We then demonstrated an application of DSRA to maximization of a pricing-based utility.
Finally, our CSRA and DSRA algorithms were compared to the state-of-the-art 
golden-section-search \cite{berry:tcom:2009} and subgradient \cite{TCOM:wong:09} based 
algorithms and shown to yield significant improvements in convergence rate.
}

\bibliographystyle{ieeetr}
\bibliography{macros_abbrev,references}

\begin{appendices}

\section{Sketch of proof for convexity of CSRA problem} \label{appendix4}
First, we show that $I_{n,k,m}F_{n,k,m}(I_{n,k,m},x_{n,k,m})$ is convex
in $I_{n,k,m}$ and $x_{n,k,m}$. For this, consider the case when
$I_{n,k,m} > 0$. In this case, the Hessian of 
$I_{n,k,m}F_{n,k,m}(I_{n,k,m},x_{n,k,m})$ w.r.t. $I_{n,k,m}$
and $x_{n,k,m}$ can be calculated and found to be positive semi-definite.
Next, consider the case when $I_{n,k,m} = 0$. To prove convexity in
this case, we apply the definition of convexity, i.e., for any two points 
$(I_{n,k,m}^{(1)}, x_{n,k,m}^{(1)})$ and $(I_{n,k,m}^{(2)},x_{n,k,m}^{(2)})$
in the domain of CSRA problem and for any $\lambda \in [0,1]$, convexity 
means
\begin{eqnarray}
\lefteqn{
\lambda I^{(1)}_{n,k,m}\,F_{n,k,m}\Big(I_{n,k,m}^{(1)}\,, x_{n,k,m}^{(1)}\Big) +
(1-\lambda)I^{(2)}_{n,k,m}\,F_{n,k,m}\Big(I_{n,k,m}^{(2)}\,, x_{n,k,m}^{(2)}\Big) 
}\nonumber \\
&\geq& \Big[\lambda I^{(1)}_{n,k,m} + (1-\lambda) I^{(2)}_{n,k,m}\Big]
F_{n,k,m}\Big(\lambda I_{n,k,m}^{(1)} + 
(1-\lambda) I_{n,k,m}^{(2)}\,, \lambda x_{n,k,m}^{(1)} + (1-\lambda) x_{n,k,m}^{(2)}\Big) .
\end{eqnarray}
When one or both of $\{I_{n,k,m}^{(1)}, I_{n,k,m}^{(2)}\}$ are zero,
it is straightforward to show that the above equation holds. 
Therefore, $I_{n,k,m}F_{n,k,m}(I_{n,k,m},x_{n,k,m})$ is convex in
$I_{n,k,m}$ and $x_{n,k,m}$. Consequently, it is a convex function of 
$\vec{I}$ and $\vec{x}$. 
Since the primal objective function of the CSRA problem
$\sum_{n,k,m}I_{n,k,m}\,F_{n,k,m}(I_{n,k,m}, x_{n,k,m})$
is a sum of functions that are convex in $\vec{I}$ and $\vec{x}$, it is also 
convex in $\vec{I}$ and $\vec{x}$.

\section{Sketch of proof of \lemref{lem3}} \label{appendix2}
Suppose that $\mu_1 < \mu_2$, where $\mu_1, \mu_2 \in [\mu_{\text{\sf min}}, 
\mu_{\text{\sf max}}]$.
With $\mu$ fixed, the minimization problem becomes
\begin{eqnarray}
\lefteqn{
L(\mu,\vec{I}^*(\mu),\vec{x}^*(\mu,\vec{I}^*(\mu)))
}\nonumber\\[-2mm]
&=& \min_{\substack{\{\vec{x} \succeq 0\} \\[0.8mm] \scriptstyle \vec{I}\in
\mc{I}\csra}} L(\mu, \vec{I},\vec{x}) 
~= \min_{\substack{\{\vec{x} \succeq 0\} \\[0.8mm] \scriptstyle
\vec{I}\in \mc{I}\csra}} \Big(\sum_{n,k,m} x_{n,k,m} -
P\con\Big)\mu + \sum_{n,k,m} I_{n,k,m} F_{n,k,m}(I_{n,k,m},x_{n,k,m}) 
\label{eq:maximin} 
\end{eqnarray}
recalling \eqref{cdual}.  
At $\mu=\mu_1$, $\vec{I}^*(\mu_2)$ and $\vec{x}^*(\mu_2,\vec{I}^*(\mu_2))$
are suboptimal values of  $\vec{I}^*(\mu)$ and 
$\vec{x}^*(\mu, \vec{I}^*(\mu))$, and at
$\mu=\mu_2$, $\vec{I}^*(\mu_1)$ and $\vec{x}^*(\mu_1,\vec{I}^*(\mu_1))$
are suboptimal values of  $\vec{I}^*(\mu)$ and 
$\vec{x}^*(\mu, \vec{I}^*(\mu))$.
Therefore,
\begin{eqnarray}
L(\mu_1,\vec{I}^*(\mu_1),\vec{x}^*(\mu_1,\vec{I}^*(\mu_1)))
&\leq& L(\mu_1,\vec{I}^*(\mu_2),\vec{x}^*(\mu_2,\vec{I}^*(\mu_2))), \label{eq:one}
~\textrm{and} \\
L(\mu_2,\vec{I}^*(\mu_2),\vec{x}^*(\mu_2,\vec{I}^*(\mu_2)))
&\leq& L(\mu_2,\vec{I}^*(\mu_1),\vec{x}^*(\mu_1,\vec{I}^*(\mu_1))). \label{eq:two}
\end{eqnarray}
Adding \eqref{one} and \eqref{two}, and evaluating the result, we get
\begin{equation}
(\mu_1 - \mu_2) \Big( \sum_{n,k,m} x_{n,k,m}^*(\mu_1,\vec{I}^*(\mu_1)) -
x_{n,k,m}^*(\mu_2,\vec{I}^*(\mu_2)) \Big) 
\leq 0.
\end{equation}
Since $\mu_1 < \mu_2$, we have
$X^*\overall(\mu_1) \geq X^*\overall(\mu_2)$.
Therefore, $X\overall^*(\mu)$ is monotonically decreasing in 
$\mu$.

\section{Proof of \lemref{lem6}} \label{appendix7}
\begin{IEEEproof}
To compare the utilities obtained by the 
proposed CSRA algorithm and the exact CSRA solution, we compare 
the Lagrangian values achieved by the two solutions. 
Recall $\mu^* \in [\underline{\mu}, \bar{\mu}] \subset 
[\mu_{\text{\sf min}}, \mu_{\text{\sf max}}]$.
Therefore,
\begin{eqnarray}
L(\mu^*, \vec{I}^*(\mu^*),\vec{x}^*(\mu^*, \vec{I}^*(\mu^*))) -
L(\underline{\mu}, \vec{I}^*(\underline{\mu}),\vec{x}^*(\underline{\mu}, 
\vec{I}^*(\underline{\mu}))) &\geq& 0, ~\textrm{and} \nonumber \\
L(\mu^*, \vec{I}^*(\mu^*),\vec{x}^*(\mu^*, \vec{I}^*(\mu^*))) -
L(\bar{\mu}, \vec{I}^*(\bar{\mu}),\vec{x}^*(\bar{\mu}, \vec{I}^*(\bar{\mu})))
&\geq& 0. \label{eq:eq5}
\end{eqnarray}
The solution of the proposed CSRA algorithm allocates resources such that 
the sum-power constraint is satisfied while achieving a Lagrangian value of 
\begin{eqnarray}
\hat{L}\csra 
&\defn& \lambda L(\bar{\mu}, \vec{I}^*(\bar{\mu}),\vec{x}^*(\bar{\mu}, 
\vec{I}^*(\bar{\mu}))) + (1-\lambda)L(\underline{\mu}, \vec{I}^*(\underline{\mu}),
\vec{x}^*(\underline{\mu}, \vec{I}^*(\underline{\mu}))). \nonumber
\end{eqnarray}
For any $\mu$, notice that
$L(\mu, \vec{I}^*(\mu), \vec{x}^*(\mu, \vec{I}^*(\mu))) = - U^*(\mu)
+ (X\overall^*(\mu) - P\con)\mu$, 
where $U^*(\mu)$ is the total utility achieved due to optimal
power allocation at that $\mu$.
Since the resource allocation obtained by the proposed CSRA algorithm and 
the exact CSRA solution satisfy the sum-power constraint with equality, we 
have
\begin{eqnarray}
U^*\csra &=& - L(\mu^*, \vec{I}^*(\mu^*),\vec{x}^*(\mu^*, \vec{I}^*(\mu^*))), \label{eq:Ucon}~\textrm{and} \\
\hat{L}\csra &=& - \hat{U}\csra(\underline{\mu}, \bar{\mu})
+ (X^*\overall(\bar{\mu}) - P\con)\lambda \bar{\mu}  + 
(X^*\overall(\underline{\mu}) - P\con)(1- \lambda) \underline{\mu} \nonumber \\
&=& - \hat{U}\csra(\underline{\mu}, \bar{\mu})
+ (X^*\overall(\bar{\mu}) - P\con)(\bar{\mu}-\underline{\mu})\lambda.
\label{eq:Ldiff}
\end{eqnarray}
Equation \eqref{Ldiff} holds since $\lambda X^*\overall(\bar{\mu}) + (1-\lambda)
X^*\overall(\underline{\mu}) = P\con$.
From \eqref{Ucon} and \eqref{Ldiff}, we get
\begin{eqnarray}
0 &\leq& U^*\csra - \hat{U}\csra(\underline{\mu}, 
\bar{\mu}) 
= - L(\mu^*, \vec{I}^*(\mu^*),\vec{x}^*(\mu^*, \vec{I}^*(\mu^*)))
+ \hat{L}\csra 
-(X^*\overall(\bar{\mu}) - P\con) (\bar{\mu}-\underline{\mu})\lambda.
\nonumber
\end{eqnarray}
From the above equation and \eqref{eq5}, we have
\begin{eqnarray}
0 ~\leq~ U^*\csra - \hat{U}\csra(\underline{\mu}, 
\bar{\mu}) &\leq& (P\con - X^*\overall(\bar{\mu}) )(\bar{\mu}-\underline{\mu}) \lambda 
\leq (\bar{\mu}-\underline{\mu})P\con.
\end{eqnarray}
\end{IEEEproof}

\section{Sketch of proof of \lemref{lem1}} \label{appendix6}

Let $\tilde{\mu} \in [\mu_{\text{\sf min}}, 
\mu_{\text{\sf max}}]$ be any value of the Lagrangian
dual variable for the CSRA problem. Then, at $\tilde{\mu}$, 
one of the following three cases holds.
\begin{enumerate}
\item $|S_n(\tilde{\mu})| \leq 1~\forall n$.
\item For some $n$, $|S_n(\tilde{\mu})| > 1$ but 
no two combinations in $S_n(\tilde{\mu})$ have the same allocated power.
\item For some $n$, $|S_n(\tilde{\mu})| > 1$ and
at least two combinations in $S_n(\tilde{\mu})$ have the same allocated power.
\end{enumerate}

We make use of two properties in the proof. Firstly, 
$V_{n,k,m}(\mu, p_{n,k,m}^*(\mu))$ is a continuous function 
of $\mu$. Therefore, by definition of continuous functions,
if $V_{n,k,m}(\tilde{\mu}, p_{n,k,m}^*(\tilde{\mu})) > 0$, then we can fix 
a $\delta_{n,k,m} \,(> 0)$ such that
$V_{n,k,m}(\mu, p_{n,k,m}^*(\mu)) > 0 
~\textrm{whenever}~ |\mu -\tilde{\mu}| < \delta_{n,k,m}$. 
Secondly, for all values of $\mu$, we know $\frac{\partial V_{n,k,m}(\mu, 
p_{n,k,m}^*(\mu))}{\partial \mu} = p_{n,k,m}^*(\mu)$.
We now apply these properties to each of the three cases to 
determine $S_n(\mu)~\forall n$.  
When $\mu$ is sufficiently close to $\tilde{\mu}$, we show that, 
in cases $1)$ and $2)$, one can fix a $\delta$ such that 
$|S_n(\mu)| \leq 1~\forall n$ whenever $0< |\mu-\tilde{\mu}| < \delta$.
When this happens, it can be shown that, for all $\mu_1,\mu_2 \in 
(\tilde{\mu} - \delta, \tilde{\mu})$, one has
$\vec{I}^*(\mu_1), \vec{I}^*(\mu_2) \,\in \{0, 1\}^{N \times 
K \times M}$ and $S_n(\mu_1) = S_n(\mu_2)~\forall n$. The same property holds 
when $\mu_1,\mu_2 \in (\tilde{\mu}, \tilde{\mu} + \delta)$. 
In case $3)$, we establish that all combinations with the same allocated 
power contribute equally to the total power allocated, as well as the total 
optimal value of Lagrangian. Therefore, all but any one combination can be 
ignored safely, implying that there exists a fixed $\delta$ such that
$\vec{I}^*(\mu) \in \{0, 1\}^{N \times K \times M}$ whenever 
$|\mu-\tilde{\mu}| < \delta$. 
After ignoring the redundant combinations, it follows from cases $1)$ and $2)$ 
that, for all $\mu_1,\mu_2 \in (\tilde{\mu} - \delta, \tilde{\mu})$ and 
$\mu_1,\mu_2 \in (\tilde{\mu}, \tilde{\mu} + \delta)$, there exists 
$\vec{I}^*(\mu_1), \vec{I}^*(\mu_2) \,\in \{0, 1\}^{N \times 
K \times M}$ such that $\vec{I}^*(\mu_1) = \vec{I}^*(\mu_2)$.


\section{Sketch of proof of \lemref{lem7}} \label{appendix8}
From \eqref{cdual} and the stated assumptions, we have
$\vec{I}^*(\mu) \in \mc{I}\dsra \subset \mc{I}\csra$ and
\begin{equation}
\Big(\vec{I}^*(\mu), \vec{x}^*(\mu, \vec{I}^*(\mu))\Big) = 
\argmin_{\substack{\scriptstyle \vec{x}\succeq 0 \\[0.8 mm]
	\scriptstyle \vec{I}\in \mc{I}\dsra}}
	\sum_{n,k,m}  I_{n,k,m}\,F_{n,k,m}(I_{n,k,m},x_{n,k,m}) + 
\Big(\sum_{n,k,m}x_{n,k,m} - P\con\Big) \mu,
\end{equation}
where $F_{n,k,m}(\cdot, \cdot)$ was defined in \eqref{defF}.
Then, applying the concept of generalized Lagrange multiplier method 
from \cite[Theorem $1$]{OR:Everett:63}, we conclude that
\begin{eqnarray}
(\mathbb{I}^*, \mathbb{X}^*) &=& \argmin_{\substack{\scriptstyle \{\mathbb{X} \succeq 0\}\\[0.8 mm]
	\scriptstyle \mathbb{I} \in \mc{I}\dsra}} \sum_{n,k,m} 
	\mathbb{I}_{n,k,m}
	F_{n,k,m}(\mathbb{I}_{n,k,m},\mathbb{X}_{n,k,m}) 
~~\text{s.t.} \sum_{n,k,m} \mathbb{X}_{n,k,m} \leq
\sum_{n,k,m} x_{n,k,m}^*(\mu, \vec{I}^*(\mu)).
\end{eqnarray}
Substituting $\mathbb{X}_{n,k,m} = \mathbb{I}_{n,k,m} 
\mathbb{P}_{n,k,m}$ back into the above equation, we obtain the desired result.

\section{Proof of \lemref{lem8}} \label{appendix9}
\begin{IEEEproof}
Let us denote $\lim_{\underline{\mu} \to \bar{\mu}} 
\hat{U}\dsra(\underline{\mu}, \bar{\mu})$
by $\hat{U}\dsra$.
The left inequality in the lemma is straightforward since 
$U^*\dsra \geq \hat{U}\dsra(\underline{\mu}, 
\bar{\mu}) ~\forall \underline{\mu}, \bar{\mu}$.
Now, if $|S_n(\mu^*)| \leq 1~\forall n$, then we have
$U^*\dsra = U^*\csra = \hat{U}\dsra$,
ensuring that the solution obtained via the proposed DSRA algorithm is optimal
in the limit $\underline{\mu}, \bar{\mu} \to \mu^*$.
However, when $|S_n(\mu^*)| > 1$ for some $n$, $P\con$
lies in one of the ``gaps'' as mentioned in \figref{sump}
and $\vec{I}^*\csra \notin \mc{I}\dsra$. 
In this case, we have
$0 \leq U^*\dsra - \hat{U}\dsra \leq U^*\csra - \hat{U}\dsra$.
Let $U^*(\vec{I})$ be the optimal utility achieved for user-MCS allocation 
matrix $\vec{I} \in \mc{I}\dsra$. We recall from 
\secref{continuous mu} that, at $\mu^*$, the allocation 
$\vec{I}^{\scriptscriptstyle\text{\sf min}}(\mu^*)$ is one of possibly many values
of $\vec{I}$ minimizing $L(\mu^*, \vec{I}, \vec{x}^*(\mu^*, \vec{I}))$. Thus, 
$U^*\csra = - L(\mu^*, \vec{I}^{\scriptscriptstyle\text{\sf min}}(\mu^*), 
\vec{x}^*(\mu^*, \vec{I}^{\scriptscriptstyle\text{\sf min}}(\mu^*)))$.
For brevity in this proof, let us denote 
$\vec{I}^{\scriptscriptstyle\text{\sf min}}(\mu^*)$ and 
$\vec{I}^{\scriptscriptstyle\text{\sf max}}(\mu^*)$ $(\in \mc{I}\dsra)$, defined
in \eqref{Imaximin}, by $\vec{I}^{\scriptscriptstyle\text{\sf min}}$
and $\vec{I}^{\scriptscriptstyle\text{\sf max}}$, respectively.
Therefore, $\hat{U}\dsra = 
\max\{U^*(\vec{I}^{\scriptscriptstyle\text{\sf min}}),
U^*(\vec{I}^{\scriptscriptstyle\text{\sf max}})\}$. This gives us
\begin{eqnarray}
U^*\csra - \hat{U}\dsra
&\leq& U^*\csra - U^*(\vec{I}^{\scriptscriptstyle\text{\sf min}}) \nonumber \\
&=& -L(\mu^*, \vec{I}^{\scriptscriptstyle\text{\sf min}}, \vec{x}^*(\mu^*, 
\vec{I}^{\scriptscriptstyle\text{\sf min}})) + 
L_{\vec{I}^{\scriptscriptstyle\text{\sf min}}}(
\mu^*_{\vec{I}^{\scriptscriptstyle\text{\sf min}}}, 
\vec{x}^*(\mu^*_{\vec{I}^{\scriptscriptstyle\text{\sf min}}})))
\nonumber \\
&=& -L(\mu^*, \vec{I}^{\scriptscriptstyle\text{\sf min}}, \vec{x}^*(\mu^*, 
\vec{I}^{\scriptscriptstyle\text{\sf min}})) + 
L(\mu^*_{\vec{I}^{\scriptscriptstyle\text{\sf min}}}, 
\vec{I}^{\scriptscriptstyle\text{\sf min}}, 
\vec{x}^*(\mu^*_{\vec{I}^{\scriptscriptstyle\text{\sf min}}}, 
\vec{I}^{\scriptscriptstyle\text{\sf min}})), \label{eq:diffL}
\end{eqnarray}
where, for \eqref{diffL}, we use the equivalence between $L(\mu, \vec{I},\vec{x})$ in \eqref{clagrange}
and $L_{\vec{I}}(\mu, \vec{x})$ in \eqref{lagrangian}. Note that
$\mu^*_{\vec{I}^{\scriptscriptstyle\text{\sf min}}} \leq \mu^*$, since the total
optimally allocated power for $\vec{I}^{\scriptscriptstyle\text{\sf min}}$ at
$\mu = \mu^*$ is less than or equal to $P\con$ and the total optimally allocated
power for any given $\vec{I}$ is a decreasing function of $\mu$. Plugging
$L(\cdot, \cdot, \cdot)$ from \eqref{clagrange} into \eqref{diffL}, we get
\begin{eqnarray}
U^*\csra - \hat{U}\dsra 
&\leq& -\Big[ -\mu^* P\con + \sum_{n,k,m}I^{\scriptscriptstyle\text{\sf min}}_{n,k,m}\Big( 
-\bar{U}_{n,k,m}(p_{n,k,m}^*(\mu^*)) + \mu^* p_{n,k,m}^*(\mu^*) \Big) \Big]  \label{eq:step}\\
& & + \Big[ -\mu^*_{\vec{I}^{\scriptscriptstyle\text{\sf min}}} P\con + \sum_{n,k,m}I^{\scriptscriptstyle\text{\sf min}}_{n,k,m}\Big( 
-\bar{U}_{n,k,m}(p_{n,k,m}^*(\mu^*_{\vec{I}^{\scriptscriptstyle\text{\sf min}}})) + \mu^*(\vec{I}^{\scriptscriptstyle\text{\sf min}}) 
p_{n,k,m}^*(\mu^*_{\vec{I}^{\scriptscriptstyle\text{\sf min}}}) \Big) \Big],\nonumber
\end{eqnarray}
where, $\bar{U}_{n,k,m}(x) = \E\big\{U_{n,k,m}\big((1-a_{k,m} e^{-b_{k,m} x \gamma_{n,k}})r_{k,m} \big)
\big\}$. Using the definition of $X\overall^*(\vec{I},\mu)$ in \eqref{Xtotal}, 
we have 
$X\overall^*(\vec{I}^{\scriptscriptstyle\text{\sf min}}, \mu^*) 
\leq P\con$ and $X\overall^*(\vec{I}^{\scriptscriptstyle\text{\sf min}}, 
\mu^*_{\vec{I}^{\scriptscriptstyle\text{\sf min}}}) = P\con$. Therefore, \eqref{step} 
can be re-written as
\begin{eqnarray}
\lefteqn{U^*\csra - \hat{U}\dsra}  \nonumber \\
&\leq& \mu^*\big( P\con - X\overall^*(\vec{I}^{\scriptscriptstyle\text{\sf min}}, \mu^*)\big) - \sum_{n,k,m}I^{\scriptscriptstyle\text{\sf min}}_{n,k,m} \Big[ 
\bar{U}_{n,k,m}(p_{n,k,m}^*(\mu^*_{\vec{I}^{\scriptscriptstyle\text{\sf min}}})) -\bar{U}_{n,k,m}(p_{n,k,m}^*(\mu^*))\Big].
\label{eq:step2}
\end{eqnarray}
Calculating the first two derivatives of $\bar{U}_{n,k,m}(x)$ with respect to $x$, 
we find that it is a strictly-increasing concave function of $x$. 
Therefore, if $x_1 \leq x_2$, one can write that
$\bar{U}_{n,k,m}(x_2) - \bar{U}_{n,k,m}(x_1) \geq 
(x_2-x_1) \bar{U}'_{n,k,m}(x_2)$.
Plugging $x_1 = p_{n,k,m}^*(\mu^*)$ and $x_2 = p_{n,k,m}^*(\mu^*_{\vec{I}^{\scriptscriptstyle\text{\sf min}}})$ into this inequality, 
we get
\begin{eqnarray}
\bar{U}_{n,k,m}(p_{n,k,m}^*(\mu^*_{\vec{I}^{\scriptscriptstyle\text{\sf min}}})) -\bar{U}_{n,k,m}(p_{n,k,m}^*(\mu^*))
&\geq& 
\Big(p_{n,k,m}^*(\mu^*_{\vec{I}^{\scriptscriptstyle\text{\sf min}}}) - p_{n,k,m}^*(\mu^*)\Big)
\frac{\partial \bar{U}_{n,k,m}(x)}{\partial x}\bigg|_{x = p_{n,k,m}^*(\mu^*_{\vec{I}^{\scriptscriptstyle\text{\sf min}}})}
 \label{eq:Udiff}
\end{eqnarray}
From \eqref{step2} and \eqref{Udiff}, we then get
\begin{eqnarray}
\lefteqn{U^*\csra - \hat{U}\dsra}  \nonumber\\
&\leq&
\mu^*\big( P\con - X\overall^*(\vec{I}^{\scriptscriptstyle\text{\sf min}}, \mu^*)\big) - 
\sum_{n,k,m} I_{n,k,m}^{\scriptscriptstyle\text{\sf min}}
\bar{U}'_{n,k,m}\big(p_{n,k,m}^*(\mu^*_{\vec{I}^{\scriptscriptstyle\text{\sf min}}})\big) 
\Big(p_{n,k,m}^*(\mu^*_{\vec{I}^{\scriptscriptstyle\text{\sf min}}}) - p_{n,k,m}^*(\mu^*)
\Big). 
\label{eq:cdineq}
\end{eqnarray}
Evaluating $\bar{U}'_{n,k,m}\big(p_{n,k,m}^*(\mu^*_{\vec{I}^{\scriptscriptstyle\text{\sf min}}})\big)$, we find
\begin{eqnarray}
\frac{\partial \bar{U}_{n,k,m}(x)}{\partial x}\bigg|_{x = 
p_{n,k,m}^*(\mu^*_{\vec{I}^{\scriptscriptstyle\text{\sf min}}})}  \hspace{-6mm}
&=&  a_{k,m} b_{k,m} r_{k,m} \E\big\{ U'_{n,k,m}\big( (1-a_{k,m} 
e^{-b_{k,m} p_{n,k,m}^*(\mu^*_{\vec{I}^{\scriptscriptstyle\text{\sf min}}})\gamma_{n,k}})r_{k,m}\big) \gamma_{n,k}e^{-b_{k,m} 
p_{n,k,m}^*(\mu^*_{\vec{I}^{\scriptscriptstyle\text{\sf min}}})
\gamma_{n,k}} \big\} \nonumber \\
&\geq& \mu_{\text{\sf min}}. \label{eq:mu_ineq}
\end{eqnarray}
From \eqref{cdineq} and \eqref{mu_ineq}, we finally obtain
\begin{eqnarray}
U^*\csra - \hat{U}\dsra
&\leq& (\mu^* - \mu_{\text{\sf min}})
\big(P\con - X\overall^*(\vec{I}^{\scriptscriptstyle\text{\sf min}}, \mu^*)\big)
~\leq~ (\mu_{\text{\sf max}} - \mu_{\text{\sf min}}) P\con .
\label{eq:step3}
\end{eqnarray}
\end{IEEEproof}

\end{appendices}

\end{document}